\DocumentMetadata{}
\documentclass[sigplan,nonacm]{acmart}

\usepackage[table]{xcolor}
\usepackage{multirow}
\usepackage{pifont}
\usepackage{tikz}
\usepackage[table]{xcolor}
\usepackage{xspace}
\usepackage{subcaption}

\newcommand{\counterevent}[1]{\texttt{#1}\xspace}
\newcommand{\DtlbLoadMissesMissCausesAWalk}{\counterevent{load.causes\_walk}}
\newcommand{\DtlbLoadMissesWalkCompletedFourk}{\counterevent{load.walk\_done\_4k}}
\newcommand{\DtlbLoadMissesWalkCompletedTwomFourm}{\counterevent{load.walk\_done\_2m}}
\newcommand{\DtlbLoadMissesWalkCompletedOneg}{\counterevent{load.walk\_done\_1g}}
\newcommand{\DtlbLoadMissesWalkCompleted}{\counterevent{load.walk\_done}}
\newcommand{\DtlbLoadMissesStlbHitFourk}{\counterevent{load.stlb\_hit\_4k}}
\newcommand{\DtlbLoadMissesStlbHitTwom}{\counterevent{load.stlb\_hit\_2m}}
\newcommand{\DtlbLoadMissesStlbHit}{\counterevent{load.stlb\_hit}}
\newcommand{\DtlbLoadMissesPdeCacheMiss}{\counterevent{load.pde\$\_miss}}
\newcommand{\DtlbStoreMissesMissCausesAWalk}{\counterevent{store.causes\_walk}}
\newcommand{\DtlbStoreMissesWalkCompletedFourk}{\counterevent{store.walk\_done\_4k}}
\newcommand{\DtlbStoreMissesWalkCompletedTwomFourm}{\counterevent{store.walk\_done\_2m}}
\newcommand{\DtlbStoreMissesWalkCompletedOneg}{\counterevent{store.walk\_done\_1g}}

\newcommand{\DtlbStoreMissesPdeCacheMiss}{\counterevent{store.pde\$\_miss}}
\newcommand{\PageWalkerLoadsDtlbLOne}{\counterevent{walk\_ref.l1}}
\newcommand{\PageWalkerLoadsDtlbLTwo}{\counterevent{walk\_ref.l2}}
\newcommand{\PageWalkerLoadsDtlbLThree}{\counterevent{walk\_ref.l3}}
\newcommand{\PageWalkerLoadsDtlbMemory}{\counterevent{walk\_ref.mem}}
\newcommand{\MemUopsRetiredStlbMissLoads}{\counterevent{load.ret\_stlb\_miss}}

\newcommand{\PageWalkerLoads}{\counterevent{walk\_ref}}

\newcommand{\Counterpoint}{CounterPoint\xspace}

\newcommand{\udd}{$\mu$DD\xspace}

\newcommand{\upath}{$\mu$path\xspace}
\newcommand{\upaths}{$\mu$paths\xspace}

\newcommand{\Upaths}{Micro-paths\xspace}

\newcommand{\uop}{$\mu$op\xspace}
\newcommand{\uops}{$\mu$ops\xspace}

\newcommand{\nodetype}[1]{\textsc{#1}}
\newcommand{\edgetype}[1]{\textsc{#1}}

\newcommand{\sig}[1]{\ensuremath {\vec{S}}(#1)} %
\newcommand{\setofpaths}[1]{\ensuremath \mathcal{P}(#1)} %

\newcommand{\defineas}{\triangleq}

\newcommand{\HPC}{HEC\xspace}
\newcommand{\HPCs}{HECs\xspace}

\def\eg{{\it e.g.}\xspace}
\def\etc{{\it etc.}\xspace}
\def\ie{{\it i.e.}\xspace}

\newcommand{\defineq}{\triangleq}

\author{Nick Lindsay}
\affiliation{
  \institution{Yale University, USA}
  \country{}
}

\author{Caroline Trippel}
\affiliation{
  \institution{Stanford University, USA}
  \country{}
}

\author{Anurag Khandelwal}
\affiliation{
  \institution{Yale University, USA}
  \country{}
}

\author{Abhishek Bhattacharjee}
\affiliation{
  \institution{Yale University, USA}
  \country{}
}

\acmConference[ASPLOS '26]{Proceedings of the 31st ACM International Conference on Architectural Support for Programming Languages and Operating Systems}{March 2026}{Pittsburgh, PA, USA}

\acmYear{2026}
\copyrightyear{2026}
\setcopyright{acmcopyright}

\acmISBN{123-4-5678-TODO-9}
\acmDOI{12.3456/TODOXXX.XXXXXXX}


\begin{document}

\title{CounterPoint: Using Hardware Event Counters to Refute and Refine Microarchitectural Assumptions}
\subtitle{Extended Version}

\begin{abstract}
\noindent
Hardware event counters offer the potential to reveal not only performance bottlenecks but also detailed microarchitectural behavior. In practice, this promise is undermined by their vague specifications, opaque designs, and multiplexing noise, making event counter data hard to interpret.

We introduce \Counterpoint, a framework that tests user-specified microarchitectural models—expressed as $\mu$path Decision Diagrams—for consistency with performance counter data. When mismatches occur, \Counterpoint pinpoints plausible microarchitectural features that could explain them, using multi-dimensional counter confidence regions to mitigate multiplexing noise. We apply \Counterpoint to the Haswell Memory Management Unit as a case study, shedding light on multiple undocumented and underdocumented microarchitectural behaviors. These include a load–store queue-side TLB prefetcher, merging page table walkers, abort\-able page table walks, and more.

Overall, \Counterpoint helps experts reconcile noisy hardware performance counter measurements with their mental model of the microarchitecture— uncovering subtle, previously hidden hardware features along the way.
\end{abstract}

\maketitle

\begingroup
\renewcommand\thefootnote{}\footnotetext{
© Nick Lindsay, Caroline Trippel, Anurag Khandelwal and Abhishek Bhattacharjee; 2026. This is the extended version of the authors' version of The Work. It is posted here for your personal use. Not for redistribution. The definitive version is to be published in Proceedings of the 31st ACM International Conference on Architectural Support for Programming Languages and Operating Systems, Volume 2 (ASPLOS '26), https://doi.org/10.1145/3779212.3790145.
}
\addtocounter{footnote}{-1}
\endgroup

\setlength{\abovedisplayskip}{3pt}   %
\setlength{\belowdisplayskip}{3pt}   %

\definecolor{tablegray}{gray}{0.9}

\section{Introduction}\label{sec:introduction}

\begin{figure*}[htbp]
    \centering
    \begin{subfigure}[t]{0.32\textwidth}
        \centering
        \includegraphics[width=0.868\linewidth]{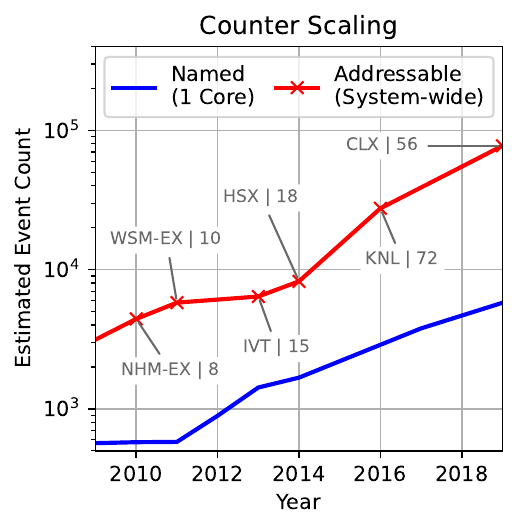}
        \caption{The estimated number of \HPCs events in x86-64 systems increased over $10\times$ between 2009 and 2019. (Y-axis plotted on a log-scale.)}
        \label{fig:EC-scaling}
    \end{subfigure}
    \hfill
    \begin{subfigure}[t]{0.32\textwidth}
    \centering
    \raisebox{0.03cm}{
        \includegraphics[width=0.9\linewidth]{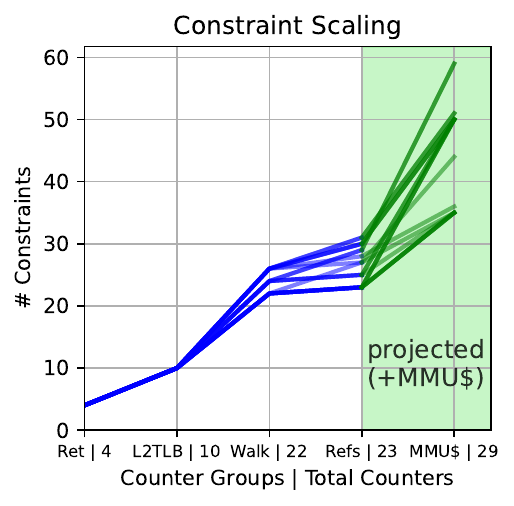}
    }
    \caption{The number of model constraints implied by a model scales with the number of \HPCs used to understand hardware behavior.
    }
    \label{fig:feasibility-constraint-scaling}
    \end{subfigure}
    \hfill
    \begin{subfigure}[t]{0.32\textwidth}
    \centering
    \includegraphics[width=0.9\linewidth]{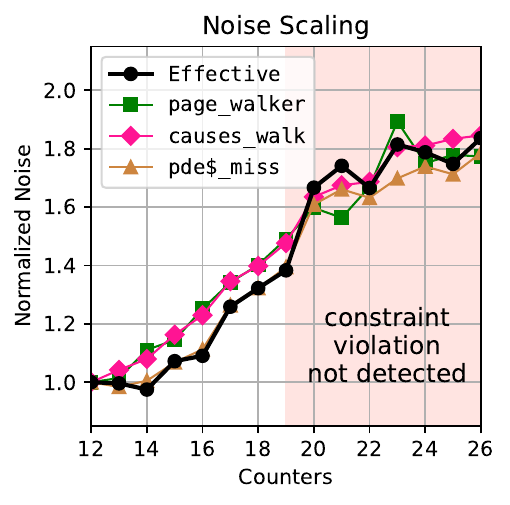}
    \caption{\HPC measurement noise increases with active \HPCs; beyond a point, model constraint violations can no longer be reliably detected.}
    \label{fig:noise-scaling}
    \end{subfigure}
    \vspace{-2mm}
    \caption[Caption]{The rapid growth of \HPCs has increased manual effort to construct and compose model constraints, and amplified multiplexing noise that obscures constraint violations. (a) The blue line shows the number of documented \HPC{} `names', assuming a single core. The red line shows the number of `addressable' events after accounting for per-core replication and the conservative removal of events that, while still documented (and potentially informative), have been deprecated by the vendor. Each red data point represents a microarchitecture paired with its typical core count in a server system. This graph shows only documented events and does not include the thousands of additional undocumented \HPCs identified in recent work \cite{ExplorationOfHiddenPmuEvents}\protect\footnotemark. (b) The number of model constraints grows superlinearly with the number of \HPCs (our x-axis shows increasing \HPC count for an Intel Haswell MMU, in steps associated with all the \HPCs in a logical group; \eg, 10 \HPCs for L2 TLB events) and worsens significantly when including hypothetical \HPCs across all MMU caches (shown in green). (c) For a representative model constraint on the Intel Haswell MMU ((1) in Table \ref{tab:feasibility-constraints}), we show that as measurement noise increases—both overall and for individual \HPCs—it becomes impossible to determine whether the model constraint is violated with 99\% confidence once 19 \HPCs are active. Here, noise is defined as the standard deviation in the observed \HPC values.}
    \vspace{-2mm}
    \label{fig:main}
\end{figure*}

Hardware event counters (\HPCs) are specialized registers embedded in CPUs and hardware accelerators that provide low-overhead, fine-grained insights into microarchitectural behavior during execution. First introduced in the 1980s—most notably in the DEC VAX and early RISC machines—\HPCs were originally designed to support performance tuning and system-level debugging.

Since then, their role has expanded. While \HPCs remain essential for identifying performance bottlenecks \cite{Yasin2014, Emer1984, lindsay2024understanding, eyerman2007top}, they are now also used to calibrate microarchitectural software simulators \cite{KanellopoulosVirtuosoMethodology, sanchez2013zsim}, build analytical models of hardware \cite{Agbarya2020,AzimiOnlineCounters,lindsay2024understanding}, correlate microarchitectural activity with power and thermal behavior \cite{isci2003runtime, lee2005using, Singh2020, Kawaguchi2016, Bircher2012, zamani2012study}, and more \cite{abel2019uops,abel2020nanobench, Park2022, zhang2007processor, azimi2009enhancing,demme2013feasibility, wang2013numchecker}. %
As their utility has grown, so has their prevalence: modern x86-64 processors now expose thousands of \HPCs—more than a 10$\times$ increase since 2009 (Figure~\ref{fig:EC-scaling}).

\vspace{1mm}
{\noindent \bf The promise of a broad set of \HPCs.} In principle, a rich set of \HPCs should allow experts to gain deeper insight into their mental model of the hardware, even without access to proprietary RTL or internal documentation. These insights are crucial for building accurate performance models and calibrating architectural simulators for future hardware \cite{bhattacharjee2019appendix, Bhattacharjee2013largereach, Bhattacharjee2018, Bhattacharjee2017, barr2010translation, Basu2013, Agbarya2020, rubin:sosp}, and go beyond traditional \HPC uses that measure only broad performance metrics like CPU and memory utilization \cite{sites_2022_perfcounters, Lv2018, pan:sosp, tang:impact, tang:optimizing, filanovsky:seeing, yang:shim}.

Address translation provides a prime example. Modern processors devote many \HPCs to this function—for instance, IBM’s Power9 includes 96 \HPCs dedicated solely to address translation \cite{IBM2018POWER9PMU}. Researchers often attempt to leverage these counters to reverse-engineer address translation hardware, enabling accurate integration into software simulators and analytical models of system performance \cite{bhattacharjee2019appendix, Bhattacharjee2013largereach, Bhattacharjee2018, Bhattacharjee2017, barr2010translation, Basu2013, Agbarya2020, rubin:sosp}. These models commonly assume specific behavior for the Paging Directory Entry (PDE) cache, which eliminates memory accesses to non-leaf page table levels during a walk. It is assumed that the PDE cache is accessed exactly once per page-table walk, implying that the number of PDE cache misses (\DtlbLoadMissesPdeCacheMiss) should not exceed page walks (\DtlbLoadMissesMissCausesAWalk):\[ \DtlbLoadMissesPdeCacheMiss \leq \DtlbLoadMissesMissCausesAWalk \]

Surprisingly, our measurements on Intel Haswell show that this expected relationship—a sanity check of the expert’s mental model, which we term a {\it model constraint}—does not always hold. This challenges a widely held assumption in address translation research and casts doubt on the validity of much simulation-based work that relies on it. %
This example also illustrates two benefits of having a diverse set of \HPCs.

First, a diverse set of \HPCs helps detect violations of the model constraints associated with an expert's mental model of the hardware. Here, we can spot the violation only because Haswell exposes the \DtlbLoadMissesMissCausesAWalk \HPC—which many other processors lack. Without it, researchers rely on generic TLB miss \HPCs that miss such nuances.

Second, a diverse set of \HPCs helps explain {\it why} a model constraint is violated, enabling refinement toward a more accurate representation of the hardware. For example, comparing counters for retired TLB misses, PDE cache misses, and page table walks uncovers two likely undocumented behaviors: (i) merged walks to the same virtual address occurring after PDE cache lookup, and (ii) aborted translation requests that terminate after PDE cache lookup but before a page table walk begins. Without the full set of \HPCs, these effects would remain hidden.

\vspace{1mm}
{\noindent \bf The reality of a broad set of \HPCs.} The PDE cache example is an ideal case where a diverse set of \HPCs helps reveal that there is a flaw in the expert's original mental model of the microarchitecture. In practice, however, experts are rarely able to leverage the full introspective power of \HPCs because they rely on manual and ad hoc approaches to doing so. In particular, they face two challenges:

\footnotetext{We counted the total number of \HPC{} names in the Linux \texttt{perf} counter database \cite{linux-pmu-events,intel-perfmon} to determine the set of `Named' events per microarchitecture. We  estimated the number of `Addressable' events by (i) conservatively removing deprecated \HPCs, (ii) distinguishing between core and uncore \HPCs, and (iii) accounting for per-core replication. We account for per-core replication by summing together the uncore counters with the number of core counters multiplied by the typical core count of server systems of that microarchitecture. We will further explore these trends in future work.}

First, understanding microarchitectural behavior requires identifying how \HPCs relate to one another—that is, determining {\it all} the model constraints that observed \HPC data must satisfy to align with an expert’s mental model (\ie, their set of assumptions about the microarchitectural implementation).
Our PDE cache model constraint illustrates how surprisingly difficult it can be to reason about even simple relationships involving just two \HPCs. As more \HPCs are used to check whether observed behavior matches expert expectations, the number of model constraints grows super-linearly (Figure \ref{fig:feasibility-constraint-scaling}). These model constraints become increasingly complex, often involving dozens of \HPCs in intricate relationships that are hard to reason about (Section \ref{sec:motivation}). Manual approaches to identifying and evaluating model constraints quickly become intractable. The challenge worsens when observed \HPC values violate model constraints, forcing experts to revise their mental models—and then deduce entirely new, complex sets of associated model constraints.

Second, modern architectures allow recording thousands of {\it logical} \HPCs, but these are multiplexed onto a much smaller number of {\it physical} \HPCs—typically just 4 to 8 at a time. This means that \HPC measurements are approximate rather than exact, leading to measurement noise that makes it even more difficult to evaluate model constraints. Multiplexing noise typically grows rapidly with the number of \HPCs being measured. Beyond a point, the growing number of \HPCs makes it nearly impossible to determine whether a representative model constraint is truly violated (Figure \ref{fig:noise-scaling}).

\begin{figure*}
    \centering
    \includegraphics[width=\linewidth]{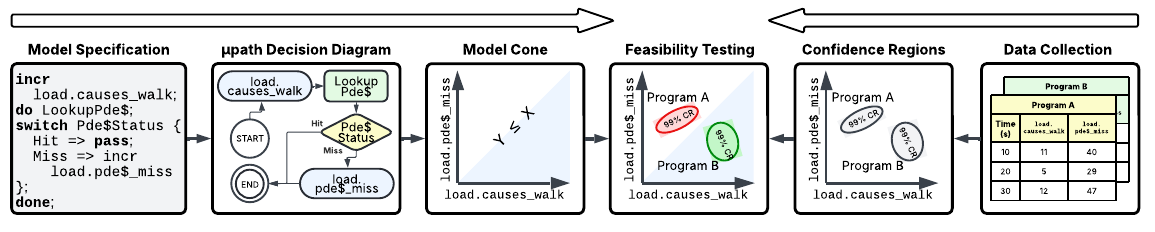}
    \vspace{-6mm}
    \caption{\Counterpoint automatically determines the feasibility of a microarchitectural model against \HPC data. Models are described using a DSL and transformed into a $\mu$path Decision Diagram ($\mu$DD), which is analyzed to determine the model cone (the set of model constraints). Counter confidence regions are constructed for each observation to handle multiplexing noise. Observations are tested against all model constraints simultaneously. \Counterpoint's counter confidence region bounds are sharper than other approaches, enabling more violations to be identified, and thereby enabling more opportunities to refine the expert's microarchitectural assumptions. \Counterpoint effortlessly supports dozens of \HPCs and constraints.}
    \vspace{-4mm}
    \label{fig:microflow-overview}
\end{figure*}

\vspace{1mm}
{\noindent \bf Extracting the promise of \HPCs with \Counterpoint.} To bridge the gap between the promise and reality of \HPCs, we invent \Counterpoint{}\footnote{CounterPoint enables using hardware event {\it counters} to {\it point} out gaps in an expert’s understanding of microarchitectures and makes it easier to explore improvements or {\it counterpoints} to their assumptions.}—a framework that helps experts reconcile \HPC data with their mental models of the microarchitecture. \Counterpoint automates the demanding task of generating all model constraints associated with a model and checking them against noisy \HPC data, enabling exploration of the accuracy of a wider range of microarchitectural models. \Counterpoint is centered around three key insights. %

First, experts can more naturally express their envisioned hardware as a directed acyclic graph (DAG) linking hardware components to \HPC activity, rather than directly constructing model constraints. DAGs are well-suited to formal tools that can automatically derive these constraints. We introduce the $\mu$path Decision Diagram ($\mu$DD)—a specialized DAG for capturing an expert’s mental model of microarchitectural structures and how interactions among these structures increment \HPCs. A $\mu$DD concisely describes a set of microarchitectural execution paths (\upaths) that micro-ops (\uops) may follow. Each \upath is associated with specific microarchitectural events, including those that increment performance counters, enabling natural and complete generation of all constraints implied by the model.

Second, experts can refine microarchitectural models more effectively when model constraints are as tightly upper- and lower-bounded as possible (\eg, constraint (3) in Table~\ref{tab:feasibility-constraints} is most useful when its left-hand side tightly lower-bounds the number of memory references in a page table walk). Tight constraints increase sensitivity to even minor deviations from the expert’s mental model. Such tightness is more likely when \HPC relationships are expressed at the granularity of micro-ops, enabling precise attribution of events to specific hardware behaviors. $\mu$DDs naturally exploit this by modeling micro-op flows through execution paths, yielding model constraints with inherently tight bounds.

Third, while more \HPCs increase multiplexing noise, they also increase intrinsic correlations (\eg, page table walks often correlate with TLB misses). These correlations allow statistical methods to build tight {\it counter confidence regions}—rang\-es of \HPC values likely to occur with a given probability from noisy data. Compared to traditional methods that treat counter noise independently \cite{Lv2018,AzimiOnlineCounters}, this approach substantially reduces the impact of noise, enabling \Counterpoint's automated analysis to scale well beyond the number of physically available \HPCs.

\vspace{1mm}
{\noindent \bf The \Counterpoint approach.} Experts begin by expressing their mental model of the microarchitecture in a domain-specific language, which \Counterpoint translates into a $\mu$DD (Figure \ref{fig:microflow-overview}). Experts also run workloads on the target hardware, collecting as many active \HPCs as needed for analysis.

Given a $\mu$DD, \Counterpoint applies convex geometry techniques to derive the \emph{model cone}—all \HPC value combinations producible by micro-ops traversing the $\mu$DD. The model cone represents the values that simultaneously satisfy all model constraints, and eliminates the need for manual derivation. \Counterpoint then processes noisy \HPC measurements from real hardware, extracting intrinsic correlations to define tight {\it counter confidence regions}: ranges of \HPC values inferred with high confidence despite multiplexing noise. %

Finally, with feasibility testing, \Counterpoint compares the counter confidence region against the model cone. If they do not intersect, the expert’s model is inconsistent with the \HPC observations, implying that some model constraints are violated. \Counterpoint reports these violations, guiding how the $\mu$DD may be revised for consistency. This enables iterative exploration: the expert proposes new $\mu$DDs, and \Counterpoint tests them until a consistent model is found.

\vspace{1mm}
{\noindent \bf Evaluating \Counterpoint via a case study.} %
We demonstrate \Counterpoint's capabilities by applying it to the Intel Haswell Memory Management Unit (MMU), where we uncover several previously undocumented and underdocumented features. These include a load–store queue-side TLB prefetcher (as well as its trigger conditions and interaction with page hotness tracking), hardware mechanisms that merge and abort page table walks, and a cache for the root level of the page table. The Haswell MMU serves as a compelling case study: it embodies complex hardware–softw\-are interactions, %
and has been foundational for a decade of address translation research \cite{park2016efficient,merrifield2016performance,doityourself,yan2017hardware,van2017reverse,ertmerreverse,Agbarya2020,Zhao2022,tatar2022tlb,manocha2022implications,lindsay2024understanding}. Yet, it is poorly modeled in state-of-the-art software simulators \cite{gem5:1, gem5:2, gem5:3, gem5:4}, motivating recent efforts to use \HPCs to reverse-engineer accurate models \cite{bhattacharjee2019appendix, Bhattacharjee2013largereach, Bhattacharjee2018, Bhattacharjee2017, barr2010translation, Basu2013, Agbarya2020, rubin:sosp}. As a rigorous test of \Counterpoint's analysis of sophisticated microarchitectural behavior, the Haswell MMU case study provides a foundation for extension to other components and more modern microarchitectures.

\vspace{1mm}
{\noindent \bf Technical contributions.} Overall, this work:

\vspace{1mm}
{\noindent -} Defines \HPC model constraints and demonstrates their ability to expose hidden microarchitectural behavior.

{\noindent -} Introduces the $\mu$DD, a compact representation that encodes both microarchitectural assumptions and \HPC semantics.

{\noindent -} Defines the model cone, shows how it can be naturally derived from the $\mu$DD, and proves its equivalence to the model constraints of the $\mu$DD.

{\noindent -} Applies counter confidence regions to mitigate measurement noise, enabling reliable inference even when the number of counters exceeds hardware limits.

{\noindent -} Develops $\mu$DD feasibility testing to automatically validate measured \HPC data against a $\mu$DD's implied constraints.

{\noindent -} Reveals several likely undocumented and underdocumented features in a commercial Intel CPU—including TLB prefetching, early paging-structure cache lookups, and merged page table walks—using \Counterpoint's automated analysis.

\vspace{1mm}
In sum, \Counterpoint uses \HPCs to refine expert understanding of hardware—challenging incorrect assumptions and uncovering subtle, otherwise hidden effects. Such insights are essential for building trustworthy models as computer systems grow increasingly complex and opaque. Because such insights are essential for building trustworthy models amid increasingly complex and opaque computer systems, we will publicly release \Counterpoint\footnote{\Counterpoint will be maintained at: \\\url{https://github.com/NicholasLindsay/counterpoint-public}}.

\begin{table*}[]
\caption{The Haswell MMU requires reasoning about dozens of model constraints; we show three representative examples. Deriving the exact constraints for a given model is challenging because they stem from subtle microarchitectural assumptions. For example, Constraint 1 relies on expert knowledge that no retired TLB miss suffered a prior page fault. Constraint 2 relies on even more subtle knowledge that an upper bound on the number of memory references injected by a page table walker is determined by (i) PDE cache hit/miss status; (ii) the page size of the translation and (iii) the fact that every walk makes at least one memory reference. 
 For brevity, we define: $\PageWalkerLoads \defineas \PageWalkerLoadsDtlbLOne +\PageWalkerLoadsDtlbLTwo + \PageWalkerLoadsDtlbLThree + \PageWalkerLoadsDtlbMemory$.}
\label{tab:feasibility-constraints}
\begin{tabular}{|p{17.3cm}|} \hline

\rowcolor{tablegray} $(1)\ \MemUopsRetiredStlbMissLoads \leq \DtlbLoadMissesWalkCompleted \hfill \textbf{2 \HPCs}$ \label{c:retired_stlb_misses} \\ \hline
Every TLB-miss micro-op that retires must have obtained a valid translation from a page-table walk. \\ \hline

\rowcolor{tablegray} $(2)\ \begin{aligned}
&\PageWalkerLoads \leq \DtlbLoadMissesMissCausesAWalk + \DtlbStoreMissesMissCausesAWalk + 3 \cdot \DtlbLoadMissesPdeCacheMiss + 3 \cdot \DtlbStoreMissesPdeCacheMiss \\
&\quad - \DtlbLoadMissesWalkCompletedTwomFourm - \DtlbStoreMissesWalkCompletedTwomFourm - 2 \cdot \DtlbLoadMissesWalkCompletedOneg - 2 \cdot
\DtlbStoreMissesWalkCompletedOneg\
\end{aligned}$ \hfill \textbf{12 \HPCs} \label{c:max_walk_length} \\ \hline
The number of memory accesses made by the page table walker is upper bounded by the distribution of combinations of page sizes and PDE cache interactions. \\ \hline

\rowcolor{tablegray} $(3)\ 
\begin{aligned}
    \DtlbLoadMissesMissCausesAWalk + \DtlbStoreMissesMissCausesAWalk + \DtlbLoadMissesWalkCompletedOneg + \DtlbStoreMissesWalkCompletedOneg \leq \PageWalkerLoads
\end{aligned}$ \hfill{\textbf{8 \HPCs}} \label{c:min_walk_length} \\ \hline
Every page table walk must result in one or more page table walker memory accesses. Walks that complete with 1GB page emit two memory references when the MMU cache for the root page table level is absent. \\ \hline

\end{tabular}

\end{table*}

\section{CounterPoint: A Bird’s-Eye Overview} \label{sec:motivation}

{\noindent \bf The pros and cons of model constraints.} Model constraints are valuable because they let experts identify exactly when and how their assumptions about the microarchitecture break down. The \HPCs involved in a violated model constraint highlight which parts of the model may be incorrect. However, to be fully effective, all (often dozens of) model constraints must be enumerated, and each must be correct and tight. By tight, we mean the bounds leave minimal slack: loose constraints can miss infeasible observations, whereas tight constraints clearly delineate what is possible versus impossible, making violations easier to detect.

Manually deriving all the constraints is onerous, even for an expert. Table \ref{tab:feasibility-constraints} shows just a subset of constraints for a simple Intel Haswell MMU model. Each may involve many \HPCs and depend on the intersection of multiple microarchitectural assumptions.

Worse, constraints are easy to formulate either too loosely or incorrectly. For example, one might bound the number of page walker loads on Haswell, with its four-level page table:
\[ \PageWalkerLoads \leq 4 \cdot (\DtlbLoadMissesMissCausesAWalk + \DtlbStoreMissesMissCausesAWalk) \]
This is correct but not tight, since it ignores page size and MMU cache hits (unlike Constraint 2 in Table \ref{tab:feasibility-constraints}).

Alternatively, one could try to exploit the fact that larger pages shorten page table walks:
\[
\begin{aligned}
    &\PageWalkerLoads \leq \\
    &\quad\quad 4\cdot\DtlbLoadMissesWalkCompletedFourk + 4\cdot\DtlbStoreMissesWalkCompletedFourk + \\
    &\quad\quad 3\cdot\DtlbLoadMissesWalkCompletedTwomFourm + 3\cdot\DtlbStoreMissesWalkCompletedTwomFourm + \\
    &\quad\quad 2\cdot\DtlbLoadMissesWalkCompletedOneg + 2\cdot\DtlbStoreMissesWalkCompletedOneg + \\
\end{aligned}
\]
But this version is too strong: it rejects valid cases where walks inject memory accesses but do not terminate (\eg, invalid translations). As Constraint 2 shows, the tightest correct bound is actually a far more nuanced relationship.

\begin{figure*}[htbp]
    \centering
    \begin{subfigure}[t]{0.24\linewidth}
    \centering
    \includegraphics[width=\linewidth]{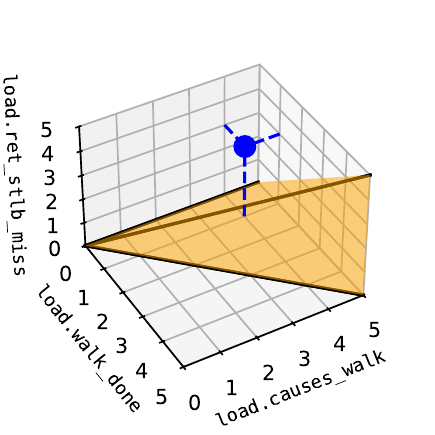}
    \caption{Candidate model cone (yellow) from three \HPCs shows a violation of a model constraint.}
    \label{fig:model-cone-3d}
    \end{subfigure}
    \hfill
    \begin{subfigure}[t]{0.24\linewidth}
    \centering
    \includegraphics[width=0.8\linewidth]{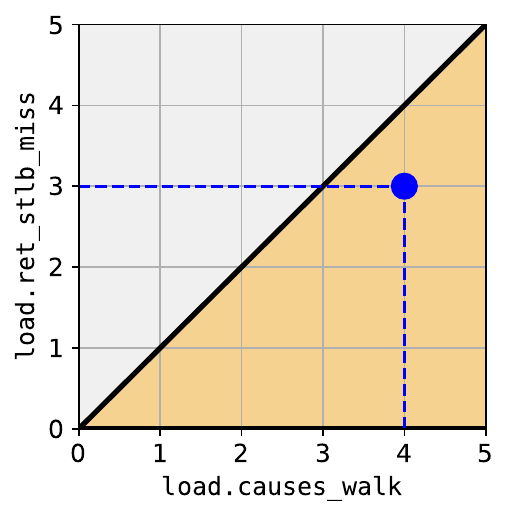}
    \caption{Ignoring an \HPC loosens the model cone and misses the violated constraint in Figure \ref{fig:model-cone-3d}.}
    \label{fig:model-cone-2d}
    \end{subfigure}
    \hfill
    \begin{subfigure}[t]{0.24\linewidth}
    \centering
    \includegraphics[width=\linewidth]{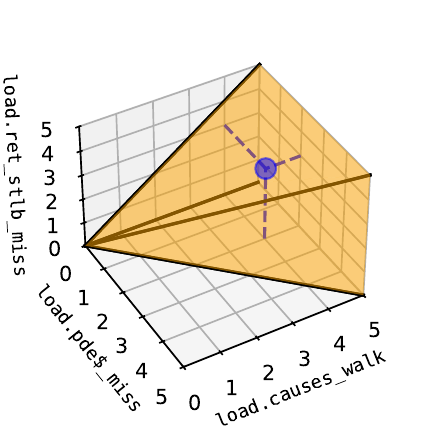}
    \caption{\HPCs with subtly different semantics loosen the cone, missing Figure \ref{fig:model-cone-3d}'s violated constraint.}
    \label{fig:model-cone-3d-alternative}
    \end{subfigure}
    \hfill
    \begin{subfigure}[t]{0.24\linewidth}
    \centering
    \includegraphics[width=0.8\linewidth]{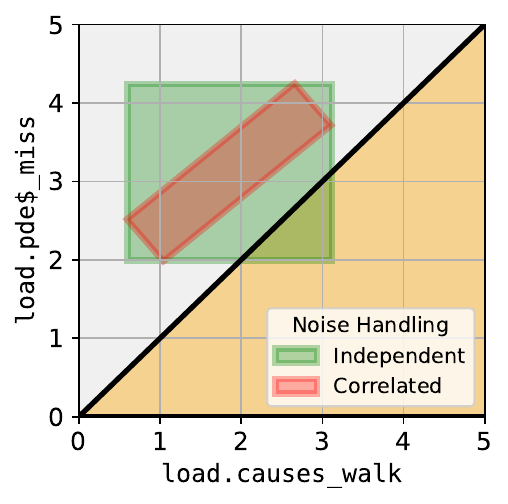}
    \caption{Counter confidence regions are tighter with \HPC correlations (red) than without (green).}
    \label{fig:confidence-region-discrimination}
    \end{subfigure}
    \vspace{-2mm}
    \caption{The ability of \HPCs to test  assumptions depends on their number and semantics, shown here pictorially. The orange regions represent points which satisfy all model constraints; the blue dot represents an observation; the red and green boxes represent two alternative constructions of counter confidence regions. Model constraints correspond to edges in 2D or faces in 3D. (a) Consider a model cone constructed from the three HECs shown. These counters imply three constraints: $\MemUopsRetiredStlbMissLoads \leq \DtlbLoadMissesWalkCompleted$ because each retired STLB miss must correspond to a unique, successfully completed page table walk; $\MemUopsRetiredStlbMissLoads \leq \DtlbLoadMissesMissCausesAWalk$ because each retired STLB miss must trigger exactly one page table walk; and $\DtlbLoadMissesWalkCompleted \leq \DtlbLoadMissesMissCausesAWalk$ because only a subset of initiated page table walks ultimately complete. The first two inequalities rely on the assumption that STLB misses are never merged. Using all three HECs clearly exposes a violation of these constraints, indicating a flaw in the expert’s mental model. (b) All three HECs were required to detect this flaw; removing $\DtlbLoadMissesWalkCompleted$ eliminates the second and third constraints, making the model violation undetectable. (c) Simply substituting $\DtlbLoadMissesWalkCompleted$ with $\DtlbLoadMissesPdeCacheMiss$ (or any other counter) is insufficient, because the semantics of each counter matter. Using this alternative counter adds the constraint $\DtlbLoadMissesPdeCacheMiss \leq \DtlbLoadMissesMissCausesAWalk$, but this constraint still fails to reveal the model violation. (d) Counter confidence regions replace point observations with value ranges; exploiting correlations yields tighter bounds than assuming independence. \\
    }
    \vspace{-2mm}
    \label{fig:model-cone-structure}
\end{figure*}

Even simpler constraints require tightness. For instance, Figure \ref{fig:model-cone-3d} shows an infeasible observation of \HPC values detectable only with enough relevant constraints. With fewer or irrelevant counters (Figures \ref{fig:model-cone-2d} and \ref{fig:model-cone-3d-alternative}), the violation slips through. When scaling to dozens of model constraints, many of which include complex relationships among dozens of \HPCs each, all these problems compound.

\vspace{1mm}
\noindent {\bf The geometry underlying model constraints.} Model constraints are powerful for validating hardware assumptions but are unscalable as they are derived in an ad hoc manner. The challenge is not in their use, but their derivation.

We observe that model constraints naturally arise because microarchitectural events occur in predefined groups rather than in isolation—for example, each completed walk for a 4KB page (\DtlbLoadMissesWalkCompletedFourk) involves 1 to 4 page table walker memory accesses (\PageWalkerLoads). Our insight is that instead of manually deriving constraints, experts can more easily enumerate all valid groupings, letting \Counterpoint automatically determine whether an observed set of events could result from some combination of groups. This approach enables scalable feasibility checking of the model constraints.

We enable experts to specify how \uops interact with the microarchitecture—including their effect on \HPCs—using $\mu$DDs. A $\mu$DD is a specialized DAG where each path represents a single \HPC group, enabling automated testing of observed \HPC values against feasibility constraints. $\mu$DDs  are centered on \uops because they form a natural unit for grouping microarchitectural events: they are familiar to experts, fine-grained enough to capture low-level hardware interactions, and directly responsible for incrementing \HPCs. The DAG representation is concise; a few nodes can efficiently describe an exponential number of \upaths.

The group-matching problem naturally induced by the $\mu$DD is fundamentally a counting problem that can be framed in terms of convex geometry. The resulting geometric object—\ie, the model cone—represents all valid combinations of \HPC values. The Minkowski–Weyl theorem from computational geometry states that every model cone has two equivalent representations: one as the set of points generated by a $\mu$DD, and the other as the set of points bounded by model constraints \cite{fukuda2004minkowskiweyl}. We leverage these dual representations by allowing the expert to express their microarchitectural assumptions in the form most natural to them—by encoding their hardware assumptions in a $\mu$DD—while enabling \Counterpoint to automatically deduce model constraints as required for user feedback.
\Counterpoint derives the model constraints using a custom algorithm which calls into an off-the-shelf \emph{convex hull} solver, as described in Section~\ref{sec:implementation}.

\vspace{1mm}
{\noindent \bf Generating tight confidence regions of \HPC observations.} Identifying flaws in the expert’s mental model requires not only a tight model cone but also computing the narrowest possible range of values that can be confidently inferred from the observed \HPC values, despite multiplexing noise.

Standard measurement tools (\eg, \texttt{perf}) report the mean and standard deviation of the samples for each \HPC, which can be used to construct counter confidence regions. Naive methods assume each \HPC is independent, resulting in overly loose counter confidence regions (Figure \ref{fig:confidence-region-discrimination}, green box) that reduce the ability to detect violations of model constraints.

Instead, we discover that \HPC values are often correlated, a finding we extract from time-series measurements. These correlations mean that the data typically have far fewer degrees of freedom than the number of counters, allowing us to construct much tighter counter confidence regions—even when dozens of counters are measured (Figure \ref{fig:confidence-region-discrimination}, red box). Tighter counter confidence regions uncover more accurate microarchitectural models.

\vspace{1mm}
{\noindent \bf Feasibility testing for guided model exploration.}
High-quality models demand detail, with $\mu$DDs describing hundreds of unique execution paths. This produces tight model constraints, but also drives rapid growth in complexity.
\Counterpoint uses linear programming to efficiently determine model feasibility, and a conic hull algorithm (Section \ref{sec:implementation}) to derive model constraints when infeasible observations occur. Violated model constraints are reported to the expert, who uses this information to formulate refined $\mu$DDs that resolve these discrepancies and represent more accurate models of the hardware.
Naturally, the precision which \Counterpoint can infer details about the microarchitectural features depends on having a dataset of \HPC observations from a rich and diverse set of programs that stress all relevant corners of the microarchitecture. Consequently, to uncover details of the Haswell MMU in our case study, we evaluated about 20 million \HPC samples across a diverse range of workloads with dozens of models, each exhibiting a unique combination of bespoke microarchitectural features. As we use \Counterpoint to refine our understanding of the hardware, we continue to expand this set of models. %

\section{From Diagrams to Model Cones} \label{sec:model-cone-construction}

\begin{figure*}[t]
  \centering
  \begin{minipage}[t]{0.47\textwidth}
      \subcaptionbox{Example $\mu$DD that describes how a \uop interacts with the STLB and PDE cache, in the presence of \DtlbLoadMissesMissCausesAWalk and \DtlbLoadMissesPdeCacheMiss \HPCs. This $\mu$DD encodes three \upaths. \label{fig:explainer-micro-flow-diagram}}{
        \includegraphics[width=\textwidth]{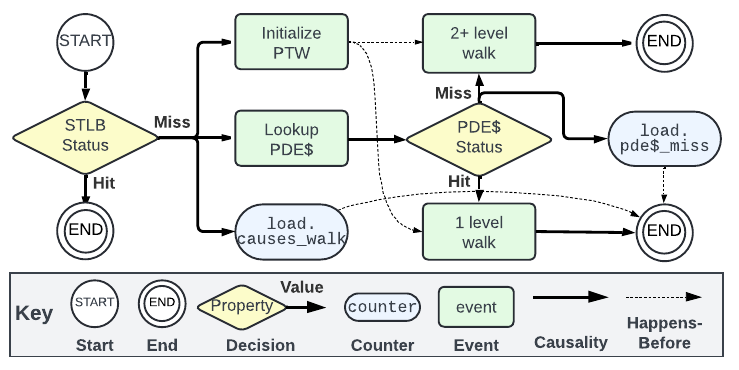}
    }
  \end{minipage}
  \hfill
  \begin{minipage}[t]{0.50\textwidth}
      \subcaptionbox{This $\mu$DD describes three unique \upaths, each corresponding to different assignments to microarchitectural properties (\eg, \emph{STLB Status} and \emph{PDE\$ Status}). Edges represent happens-before order. \label{fig:explainer-upaths}} {
        \includegraphics[width=\textwidth]{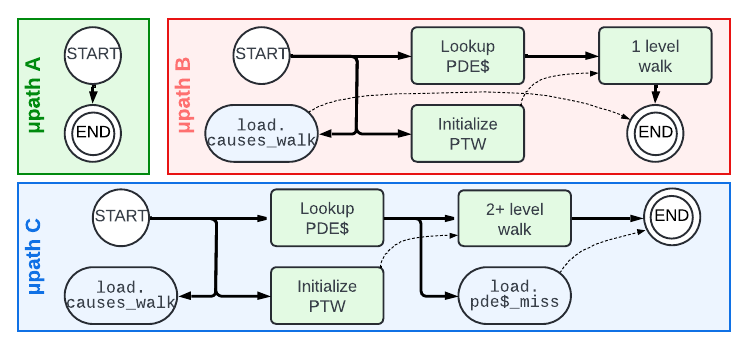}
      }
      \label{fig:combined}
  \end{minipage}
  \vspace{-2mm}
  \caption{A $\mu$DD encodes a set of microarchitectural execution paths (\upaths). Each \upath describes a set of events per \uop.}
  \label{fig:explainer-mfd-and-upaths}
\end{figure*}

{\noindent \bf $\mu$path decision diagrams.} A $\mu$DD encodes a set of microarchitectural execution paths that individual \uops may take through part of the microarchitecture (Figure \ref{fig:explainer-micro-flow-diagram} is an example encompassing a subset of address translation hardware). At the core of \Counterpoint is the concept of a microarchitectural execution path, or \upath: a happens-before ordered set of hardware events induced
by a \uop~\cite{Hsiao2024}.

\Upaths are derived from a $\mu$DD by performing a graph search along \edgetype{causality} edges, with nodes and \edgetype{causality} edges added to the \upath as they are encountered.
When a \nodetype{decision} node is encountered, there are two possibilities.
If the property has been assigned a value (determined by the labels on outgoing edges) earlier in the traversal, then the corresponding outgoing \edgetype{causality} edge is followed. %
Otherwise, a concrete property value from the outgoing \edgetype{causality} edge labels is selected and the corresponding edge is followed.
This  process continues until all nodes and \edgetype{causality} edges have been added.
\edgetype{happens-before} edges between node pairs are instantiated in the \upath if there exists a \edgetype{happens-before} edge between the corresponding $\mu$DD nodes.

When exhibiting a \upath during its execution, a \uop generates
events in a time order that respects both \edgetype{causality} and \edgetype{happens-before} edges. Events come in two forms:
standard \nodetype{event} nodes (green boxes), which represent standard microarchitectural events, and \nodetype{counter} nodes (blue pills), which correspond to events directly recorded by \HPCs.

\vspace{1mm}
\noindent {\bf $\mu$path counter signatures.}
Each \upath has an associated counter signature—a vector that records how many times each \HPC appears within a \upath. This signature captures how a \uop following that \upath increments the \HPCs.

\vspace{1mm}
{\noindent \bf Counter flow equation.} Our first goal is to precisely define when an observed set of \HPC values is “feasible” with respect to a $\mu$DD. We do so via the \emph{counter flow equation}, which links \HPC values to the number of \uops traversing microarchitectural execution paths.

The key insight behind the counter flow equation is that \uops increment \HPCs as they traverse a \upath, creating a direct relationship between the microarchitectural \emph{flow} of \uops and the resulting \HPC values.

Let $\mathcal{D}$ be a micro flow diagram and $\setofpaths{D}$ the set of \upaths it encodes.
A microarchitectural \emph{flow} $f(\cdot)$ assigns each \upath $p \in \setofpaths{D}$ a non-negative number of \uops traversing it.
Each \uop on \upath $p$ increments the \HPCs according to the \upath’s
\emph{counter signature} $\sig{p}$—the vector of counter occurrences along $p$.
Thus, the contribution of \upath $p$ is $f(p) \cdot \sig{p}$, and the total \HPC value vector $\vec{v}$ is the sum over all \upaths:
\begin{equation} \tag{Counter Flow Equation} \label{eqn:cfe}
\vec{v} = \sum_{p \in \setofpaths{D}} \sig{p} \cdot f(p)
\end{equation}
This \emph{counter flow equation} links observed \HPC values to the flow of \uops through the $\mu$DD, and is only valid when $f(p) \geq 0$ for all \upaths, as negative flows of \uops are impossible.
Intuitively, the final counter values are given by the total number of \HPC increments across all dynamic \uop instances.

\begin{figure*}[htbp]
    \centering
    \begin{subfigure}[b]{0.32\textwidth}
    \centering
    \includegraphics[width=0.76\linewidth]{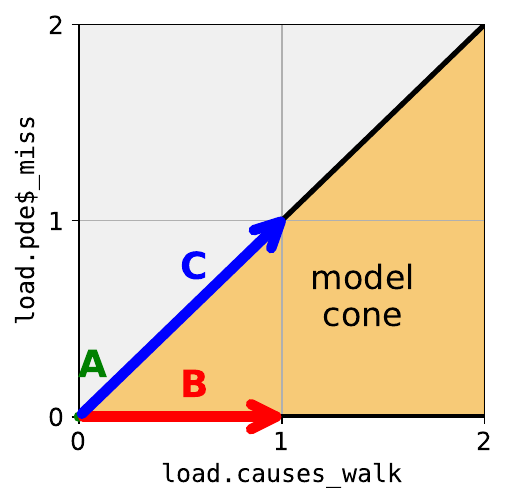}
    \caption{The model cone (valid \HPC combinations) is determined by \upath counter signatures.}
    \label{fig:model-cone}
    \end{subfigure}
    \hfill
    \begin{subfigure}[b]{0.32\textwidth}
    \centering
    \includegraphics[width=0.84\linewidth]{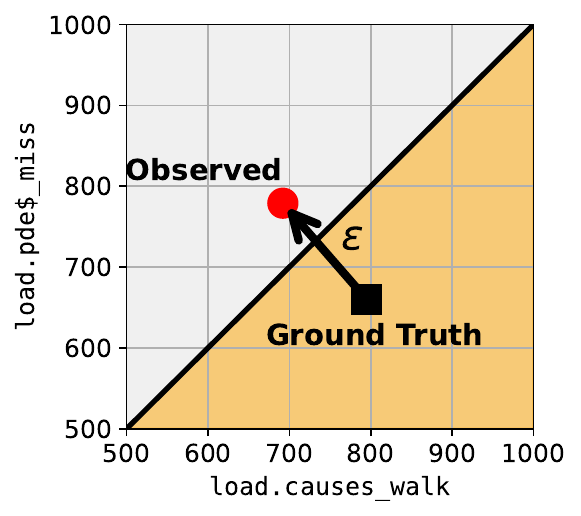}
    \caption{Noise introduced by multiplexing can make valid combinations appear infeasible.}
    \label{fig:model-cone-noise}
    \end{subfigure}
    \hfill
    \begin{subfigure}[b]{0.32\textwidth}
    \centering
    \includegraphics[width=0.79\linewidth]{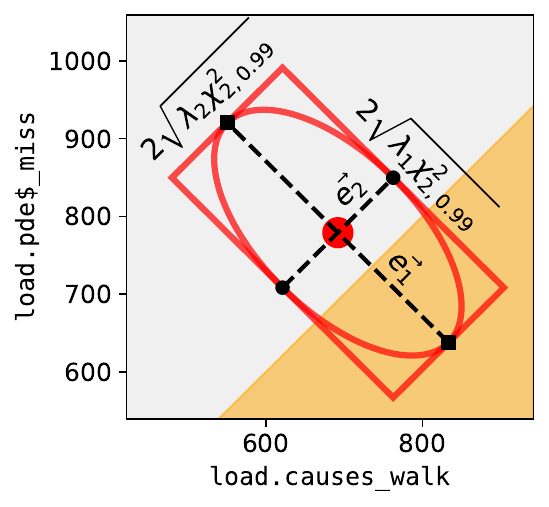}
    \caption{Confidence region construction oriented along eigenvectors of covariance matrix.}
    \label{fig:confidence-region-bounding-box}
    \end{subfigure}
    \vspace{-2mm}
    \caption{The model cone is determined purely by the \upath counter signatures (Figure a). Testing observations for inclusion in the model cone is complicated by noise, which can cause observations to spuriously appear infeasible (Figure b). \Counterpoint handles noise by constructing confidence regions at the 99\% confidence level (Figure c). The counter confidence region is an ellipsoid which \Counterpoint approximates by its bounding box, enabling a linear programming formulation. The scale and orientation of the confidence region is determined by (i) the confidence level and (ii) correlations in the observed data. $\lambda_k$ and $\vec{e_k}$ denote the $k$th eigenvalue/eigenvector of the estimated covariance matrix.}
    \label{fig:geometry}
\end{figure*}

\vspace{1mm}
{\noindent \bf Deriving the model cone.} %
The model cone is the set of all \HPC value combinations generated by valid microarchitectural executions (\ie, those with \emph{non-negative flows}).
We determine observation feasibility by testing if it lies within the model cone, a task accomplished using linear programming. This means that feasibility can be determined even without knowing $f(\cdot)$ exactly.

Mathematically, we define a model cone $K_{\mathcal{D}}$ for a $\mu$DD $\mathcal{D}$ as the set of \HPC values that are generated by microarchitectural executions with non-negative flow:
\begin{equation} \tag{Model Cone} \label{eqn:model-cone}
K_{\mathcal{D}} \defineq \left\{  \sum_{p \in \setofpaths{D}} \sig{p} \cdot f(p) \ \middle|\  f(p) \geq 0 \right\}
\end{equation}
Geometrically, $K_{\mathcal{D}}$ is a convex\footnote{\textbf{Convex}: If $x,y \in K_{\mathcal{D}}$ then $\alpha x + (1-\alpha)y \in K_{\mathcal{D}}$ for $0 \leq \alpha \leq 1$.} polyhedral\footnote{\textbf{Polyhedral}: Defined by a \emph{finite} number of equalities and inequalities.} cone\footnote{\textbf{Cone}: If $x \in K_{\mathcal{D}}$ then $\alpha x \in K_{\mathcal{D}}$ for $\alpha > 0$.} defined purely by the \upath counter signatures in the $\mu$DD (Figure \ref{fig:model-cone}).
Intuitively, the model cone represents the space of all allowed \HPC combinations.

\vspace{1mm}
\noindent {\bf Generalizability.} Our decision to design \Counterpoint so that it links fine-grained microarchitectural events and interactions—represented as $\mu$paths of $\mu$ops—with HEC updates is intentional. $\mu$op-centered execution paths have been used extensively in prior work to model low-level hardware, including formal verification of memory consistency and its interactions with coherence and virtual memory \cite{lustig2014pipecheck,Lustig2016,Manerkar2015,Hsiao2021SynthesizingImplementations}, side-channel security \cite{Trippel2018,Hsiao2024}, and functional correctness \cite{lustig2014pipecheck,Hsiao2024}. Because these approaches cover many aspects of CPU pipelines, they suggest that \Counterpoint is well positioned to extend to other microarchitectural components.

\section{Feasibility Testing with Noise}\label{sec:feasibility-testing-with-noise}

An observation is feasible if it resides within the model cone; a problem solvable with linear programming.
Unfortunately, observations are subject to multiplexing noise which must be accounted for to prevent false violations (\eg, Figure \ref{fig:model-cone-noise}).

\vspace{1mm}
\noindent {\bf Handling noise with counter confidence regions.}
Counter confidence regions handle noise by treating each observation not as a single value, but instead as a point drawn from a set of values within which the true value is likely to occur, given the presence of noise in the measurement. The likelihood of the region capturing the true value is given by the confidence level, fixed
to 99\% for our analyses. The size and shape of the counter confidence region depends on parameters of the underlying distribution, which can be inferred from the \HPC measurements themselves. \Counterpoint computes covariances (in addition to means and variances computable by \text{perf}), producing tight counter confidence regions that are more likely to catch violated constraints.

\Counterpoint requires \HPC vector samples $\{ Y_i \}_{i=1}^M$ recor\-ded at regular time intervals (\eg, every 10 seconds) over the course of a program’s execution. Such functionality is provided by standard tools (\eg, \texttt{perf}). \Counterpoint computes the sample mean $\bar{Y}$ as a \HPC vector representative of the entire execution. Statistically, the sample mean is drawn from a Gaussian distribution per the Central Limit Theorem.
With Gaussian distributions, the confidence region is fully determined by the sample mean and sample mean covariance \cite{psu_stat505_chapter4}. We calculate the \HPC covariance matrix $\Sigma_{Y}$. We estimate the sample mean covariance with the plugin estimator $\Sigma_{\bar{Y}} = \frac{1}{M} \Sigma_{Y}$. This defines the confidence region:
\begin{equation} \tag{Confidence Ellipsoid} \label{eqn:confidence-ellipse}
    \left\{\ \vec{v}\ \middle|\    (\vec{v} - \bar{Y})^T \Sigma_{\bar{Y}} (\vec{v} - \bar{Y}) \leq \chi_{N,\alpha}^2\  \right\}
\end{equation}
Intuitively, this means that the confidence region is an ellipsoid in shape, and that the ground truth (\eg, noise-free) counter value is contained within the ellipsoid with $(1-\alpha)$-confidence.
We adapt it to a linear program in the following section.
The confidence region can be made tighter by obtaining more samples (\eg, with longer running programs), providing the program has consistent steady-state behavior.

\vspace{1mm}
\noindent {\bf Determining feasibility with a linear program.} Given a model cone and a counter confidence region, we can assess the feasibility of an \HPC observation at a specified confidence level. %
If the counter confidence region intersects the model cone, the observation is deemed feasible. If there is no intersection, the observed \HPC values must violate at least one model constraint at that confidence level.
For example, Figure \ref{fig:confidence-region-bounding-box} shows a counter confidence region which intersects with the model cone, indicating a feasible observation.

To test for feasibility, \Counterpoint uses linear programming because of its efficiency, relative simplicity, and availability in mature software libraries \cite{mitchell2011pulp,diamond2016cvxpy,agrawal2018rewriting,john_forrest_2024_13347261}. \Counterpoint constructs a linear program\footnote{The full linear program is provided in Appendix \ref{apx:linear-program}.} by instantiating non-negative variables for the flow and counter values (see Section \ref{sec:model-cone-construction}).
The flows and counter values are related by the \ref{eqn:cfe}, implicitly describing the model cone. The counter confidence region, being a quadratic form, cannot be directly encoded. Instead, we approximate it with a bounding hyper-rectangle (Figure \ref{fig:confidence-region-bounding-box}).
This bounding box is aligned with the principal components of the data, producing the tightest rectangular bound on the confidence region. Empirically, our bounding box approximation detects many surprising constraint violations (Section \ref{sec:case_study}).
We leave alternatives like quadratic programming for future work.

\section{Guided Model Exploration} \label{sec:guided-model-exploration}

\noindent {\bf Discovering microarchitectural features.} The specific model constraints that are violated guide the expert in identifying which microarchitectural features need to be added or modified in the $\mu$DD to make it feasible. When a model constraint of the form
$a \cdot x \leq b \cdot x$ is violated, then for all feasible $\mu$DDs, there must exist a \upath whose $\mu$path counter signature $\sig{p}$ satisfies $a \cdot \sig{p} > b \cdot \sig{p}$.

\begin{figure}
    \centering
    \begin{subfigure}[t]{\linewidth}
        \centering
        \includegraphics[width=0.8\linewidth]{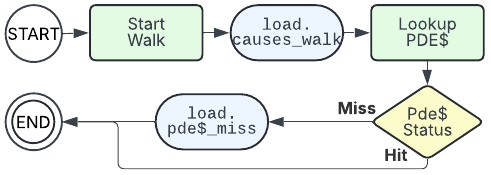}
        \caption{Initial model.}
        \label{fig:model-refinement-before}
    \end{subfigure}
    \begin{subfigure}[t]{\linewidth}
        \centering
        \vspace{0.5em}
        $\mathcal{C} \defineas \DtlbLoadMissesPdeCacheMiss \leq \DtlbLoadMissesMissCausesAWalk$
        \caption{Violated model constraint.}
        \label{fig:model-refinement-constraint}
    \end{subfigure}
    \begin{subfigure}[t]{\linewidth}
        \centering
        \includegraphics[width=0.8\linewidth]{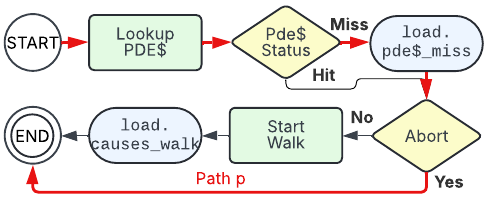}
        \caption{Refined model.}
        \label{fig:model-refinement-after}
    \end{subfigure}
    \begin{subfigure}[t]{\linewidth}
        \centering
        \vspace{0.5em}
\begin{tabular}{|c|c|c|c|c|}
\hline
\rowcolor{tablegray}
\textbf{\begin{tabular}[c]{@{}c@{}}Pde\$\\ Status\end{tabular}} &
\textbf{Abort} &
\textbf{\counterevent{\begin{tabular}[c]{@{}c@{}}load.\\ causes\_walk\end{tabular}}} &
\textbf{\counterevent{\begin{tabular}[c]{@{}c@{}}load.\\ pde\$\_miss\end{tabular}}} &
\textbf{\begin{tabular}[c]{@{}c@{}}Satisfies\\ $\mathcal{C}$\end{tabular}} \\ \hline
\textcolor{red}{Miss} & \textcolor{red}{Yes} & \textcolor{red}{0} & \textcolor{red}{1} & \textcolor{red}{No} \\ \hline
\end{tabular}
        \caption{Properties of \upath $p$ including $\mu$path counter signature.}
        \label{fig:model-refinement-new-signature}
    \end{subfigure}
    \vspace{-4mm}
    \caption{Modifying $\mu$DD to remove constraint violations is equivalent to identifying candidate microarchitectural features. (a) Initial $\mu$DD of page table walk. (b) Model implies this model constraint, which is violated. (c) $\mu$DD is updated by (i) assuming PDE cache is looked up prior to starting walk, and (ii) allowing translation requests to be aborted before starting a walk. (d) Model no longer implies constraint $\mathcal{C}$ as $\sig{p}$ does not satisfy constraint.}
    \label{fig:model-refinement}
\end{figure}

We illustrate this with an example (Figure \ref{fig:model-refinement}).
Figure \ref{fig:model-refinement-before} is a simple $\mu$DD for load \uops upon a TLB miss.
We assume that the load \uop first initializes the page table walker - incrementing \DtlbLoadMissesMissCausesAWalk - before looking up the PDE cache.
In the event of a cache miss, \DtlbLoadMissesPdeCacheMiss is incremented.
This model implies model constraint $\mathcal{C}$ (\ref{fig:model-refinement-constraint}).

\Counterpoint identifies that Constraint $\mathcal{C}$ is violated by one or more \HPC observations\footnote{When an observation is deemed infeasible with respect to an $\mu$DD, \Counterpoint automatically tests the observation against each feasibility constraint to identify violations. Deriving and testing the feasibility constraints is a non-trival procedure; we describe our implementation in Section \ref{sec:implementation}.}.
Therefore there are workloads where \DtlbLoadMissesPdeCacheMiss exceeds \DtlbLoadMissesMissCausesAWalk.
To explain this apparent contradiction, we must introduce one or more microarchitectural features into the $\mu$DD that allow for this constraint to be broken.
This corresponds to modifying the $\mu$DD such that it contains \upath{(s)} whose $\mu$path counter signatures explicitly violate $\mathcal{C}$.

One way to resolve this is to assume that (i) the PDE cache is accessed before starting a page table walk, and (ii) translation requests can be aborted between the PDE cache lookup and the start of the walk. This allows lookups to access the PDE cache without incrementing \DtlbLoadMissesMissCausesAWalk. Applying these assumptions produces a new $\mu$DD (Figure \ref{fig:model-refinement-after}), with a new \upath $p$ whose $\mu$path counter signature $\sig{p}$ explicitly violates constraint $\mathcal{C}$ (Figure \ref{fig:model-refinement-new-signature}). Analysis of this $\mu$DD confirms that $\mathcal{C}$ is no longer implied, resolving the violation. In practice, many constraints often need resolution, requiring an iterative $\mu$DD refinement process.

\begin{figure}
    \centering
    \includegraphics[width=0.55\linewidth]{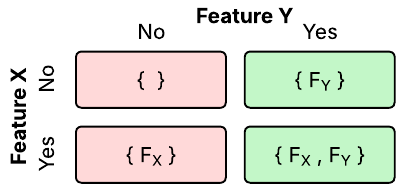}
    \vspace{-4mm}
    \caption{Microarchitectural models (boxes) classified by their features and consistency with hardware performance counter data. Red: inconsistent. Green: consistent.}
    \label{fig:microarchitectural-model-consistency}
\end{figure}

\vspace{1mm}
\noindent {\bf Classifying microarchitectural models.} Feasibility testing partitions the set of $\mu$DD into subsets of feasible and infeasible $\mu$DDs. It is possible for different $\mu$DDs representing different microarchitectural assumptions to be feasible.
When this happens, experts can identify common structures in feasible $\mu$DDs to determine likely hardware features despite the ambiguity introduced by multiple feasible $\mu$DDs.

Consider Figure \ref{fig:microarchitectural-model-consistency}.
There are four models, each identified by the presence or absence of features $F_X$ and $F_Y$.
Consistent models are highlighted in green and inconsistent models in red.
There are two consistent models; introducing ambiguity in what model is the best fit.
However, \emph{all} consistent models contain Feature $F_Y$. If we have covered the relevant feature space—by running a wide enough range of programs to ensure that the hardware we are trying to reverse-engineer is adequately exercised—then \Counterpoint can reliably conclude that Feature $F_Y$ \emph{must} be present. %
On the other hand, Feature $F_X$ in isolation is insufficient to explain the performance counter observations, but it \emph{is} possible that Feature $F_X$ and Feature $F_Y$ are both present.
Given a feature space (\eg, $F_X$ $\times$ $F_Y$), we can infer viable feature combinations.

\begin{figure}
    \centering
    \includegraphics[width=\linewidth]{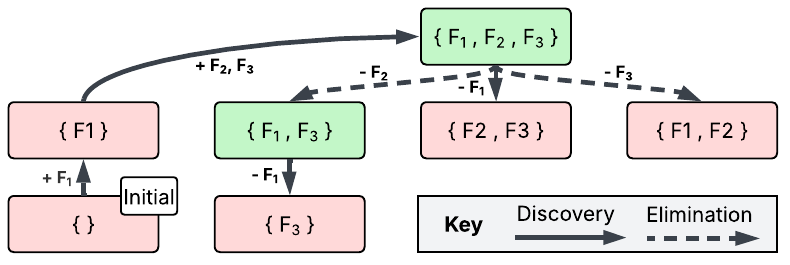}
    \vspace{-6mm}
    \caption{Expert-in-the-loop heuristic search algorithm navigates the model space without needing to explore the full cross-product of microarchitectural features. $F_\cdot$ denote microarchitectural features; models (boxes) are identified by (i) their set of features and (ii) their consistency with \HPCs.}
    \label{fig:model-search-algorithm}
\end{figure}

\vspace{1mm}
\noindent {\bf Enumerating models.}
Feature discovery and model classification can be employed together to infer the presence of microarchitectural features.
We propose a expert-in-the-loop algorithm for this purpose.

Our algorithm accepts an initial $\mu$DD and a dataset of \HPC observations, and returns a set of $\mu$DD characterized by their features and feasibility.
The algorithm consists of two phases: \emph{discovery} and \emph{elimination}.
We advocate starting with a conservative model to ensure a more informative set of constraints that enable discrimination between candidate features, but the expert can start with any model.
Features are discovered through the \emph{Discovery} phase.
Figure \ref{fig:model-search-algorithm} shows an example search graph generated by the algorithm.

\vspace{1mm}
{\noindent \it Discovery phase.} %
Constraint violations are detected by \Counterpoint, and the expert user eliminates the constraints by introducing new microarchitectural features or modifying existing ones.
When more than one feature can eliminate a constraint, all features should be added to their model.
This process is repeated until a feasible $\mu$DD is obtained.

In Figure \ref{fig:model-search-algorithm}, the initial $\mu$DD is shown at the bottom left.
Feature $F_1$ is added to produce a new model $\{ F_1 \}$; features $F_2$ and $F_3$ are then added to create the $\mu$DD at the top of the tree $\{ F_1, F_2, F_3 \}$.
This $\mu$DD is a candidate microarchitectural model for the system.
At each iteration step, the model cones are verified to ensure that the model cone is expanded.

\vspace{1mm}
\noindent {\it Elimination phase.} The candidate $\mu$DD may contain more features than required for a feasible model. In the elimination phase, we recommend recursively pruning microarchitectural features until infeasible $\mu$DDs are obtained.
This is based on our empirical observation--captured through refinement of close to a hundred models--that pruning infeasible models tends to produce infeasible models, so the sub-tree need not be explored further.
In Figure \ref{fig:model-search-algorithm}, features $F_1$-$F_3$ are removed from the top $\mu$DD to create separate $\mu$DDs.
The $\mu$DD $\{ F_1, F_3\}$ remains feasible, so features $F_1$ and $F_3$ are removed separately, resulting in infeasible $\mu$DDs.

\section{Implementing CounterPoint} \label{sec:implementation}

We implement \Counterpoint as a Python library with roughly 3K lines of code, integrating with Pandas \cite{the_pandas_development_team_2025_15831829} for convenient data processing. To support broader community adoption, \Counterpoint is designed for easy portability using a reproducible Docker \cite{merkel2014docker} environment. We will share our MMU $\mu$DDs to help seed the development of improved MMU models in widely used software simulators \cite{gem5:1, gem5:2, gem5:3, gem5:4}.

\vspace{1mm}
{\noindent \bf Domain-specific language for $\mu$DDs.} We introduce
a simple DSL for specifying $\mu$DDs: \nodetype{action} and \nodetype{counter} nodes are single-line statements, \nodetype{done} nodes use the \texttt{done} keyword, and \nodetype{decision} nodes are expressed with C-style switch cases. The DSL does not support functions, loops, or variables beyond \upath properties. Our DSL serves as a reference implementation, avoiding errors that could arise from deriving $\mu$DDs directly from RTL or C/C++ simulator specifications. %

\vspace{1mm}
{\noindent \bf Feasibility testing.} Given a $\mu$DD and a set of \HPC observations, \Counterpoint tests each observation for feasibility by constructing and solving a linear program (Appendix \ref{apx:linear-program}). This entails enumerating every counter and $\mu$path counter signature, implemented by a breadth-first traversal of the $\mu$DD. The back-end LP solver we use is \texttt{pulp} \cite{mitchell2011pulp}.
Constraint violations are identified by testing infeasible observations against the half-space defined by each constraint.

\vspace{1mm}
{\noindent \bf Deducing model constraints.} Model constraints are derived from $\mu$path counter signatures as follows.
First, $\mu$path counter signatures are normalized by dividing each element by the greatest common factor, and duplicates are removed.
Second, Gaussian elimination identifies equality constraints and eliminates redundant \HPCs\footnote{For example, consider the following relationship:\\ $\DtlbLoadMissesStlbHit = \DtlbLoadMissesStlbHitFourk + \DtlbLoadMissesStlbHitTwom$.}.
Third, \upath counter signatures that lie fully within the interior of the model cone are identified using linear programming and removed.
Fourth, the \emph{conic hull} is computed by: (i) adding the zero vector to the set of \upath counter signatures; (ii) computing the \emph{convex hull}; (iii) selecting all faces which contain the origin, corresponding to the faces of the cone.
The inequality model constraints are given by the planar equations of the resulting faces.
We implemented this custom solution because no Python library computes conic hulls, and standard numeric methods (\eg, QR factorization) are ill-conditioned, whilst symbolic operations preserve exact integer values.

\section{A Case Study: The Intel Haswell MMU} \label{sec:case_study}

{We demonstrate \Counterpoint's capabilities and evaluate its usability and performance on the Intel Haswell MMU. This case study shows how \Counterpoint can uncover the behavior of advanced microarchitectural components, even when they interact deeply with complex systems software. %
The Haswell MMU is a strong case study target due to its rich set of \HPCs \cite{intel-perfmon,linux-pmu-events} and its frequent use in prior research on address translation \cite{yan2017hardware,Agbarya2020,lindsay2024understanding,cox2017efficient,doityourself,panwar2019hawkeye}. Haswell also exhibits complex microarchitectural interactions across data and instruction activity \cite{10.1145/3669940.3707247}, as well as under native and virtualized execution \cite{bhargava2008accelerating,gandhi2016agile,adams2006comparison,ahn2012revisiting,bhatia2009performance,chang2013improving,panwar2021fast,yan2017hardware,pham2015large}. For these reasons, validating and/or refuting assumptions about the Haswell MMU represents a strong test of \Counterpoint's effectiveness. This foundation positions us to extend our study to more modern architectures. For this study, we focus specifically on data-side activity in native execution. While full confirmation of our findings would require proprietary RTL, \Counterpoint enables high confidence conclusions possible even without direct access to the RTL.

\vspace{-2mm}
\subsection{Guided Model Exploration.}
Our initial model of the Intel Haswell MMU includes features that are well-established through documentation and prior research \cite{Bhattacharjee2013largereach,Pham2014,cox2017efficient,Pham,Gosakan2023MosaicPages}, and are typically integrated in software simulators.
We assume a two-level TLB hierarchy and a four-level page table.
Building on reverse-engineering studies of Haswell MMU caches \cite{van2017reverse,ertmerreverse}, we assume the presence of a PDE cache and an additional MMU cache for the page table level immediately preceding the PDE level. Consistent with conventional wisdom, we further assumed that the PDE cache is consulted once during every walk.

We refine the model using a diverse set of \HPC observations from workloads that stress the MMU.
We measured workloads from the GAPBS \cite{beamer2017gapbenchmarksuite}, SPEC2006 \cite{henning2006spec}, PARSEC \cite{bienia2008parsec}, and YCSB \cite{cooper2010benchmarking} benchmark suites, sweeping memory footprints from 250 MB to 600 GB using input generators. We also collected \HPC data for two microbenchmarks: a linear access pattern (parametrized by footprint, stride, and load-store ratio) and a random access pattern (parametrized by footprint and load-store ratio). Through ablation studies, we found that removing these microbenchmarks causes us to miss violations of key model constraints (e.g., Constraint (1) in Table \ref{tab:feasibility-constraints}) that are essential for reverse-engineering the presence and trigger conditions of the TLB prefetchers described below\footnote{We ensured that all of our HEC measurements were unaffected by any published HEC errata. For errata that are triggered when SMT is enabled (\eg, HSD29/HSM30 affecting \texttt{mem\_uops\_retired}), we addressed this by disabling SMT in the BIOS.}. To stress different MMU behaviors, experiments were repeated with 4 KB, 2 MB, and 1 GB page sizes. Together, these workloads and configuration options yield about 20 million \HPC samples—enough observations to thoroughly stress-test our model assumptions and drive higher-quality model refinement.

We evaluate our models at the 99\% confidence level. Across dozens of representative $\mu$DDs, we found that correlated counter confidence regions detect over 24\% more model constraint violations compared to confidence regions that assume \HPCs are independent. For some models, exploiting correlations revealed over 75\% additional violations compared to baseline. \Counterpoint's confidence regions are effective because \HPCs are highly correlated: in our dataset we find that over 25\% of counter pairs have a Pearson correlation coefficient that exceeds $0.9$ (where $1.0$ indicates perfect correlation, and $0.0$ indicates no correlation). %

With \Counterpoint's support for guided model refinement, we explored dozens of $\mu$DDs.
Our initial $\mu$DD contained 31 constraints, 8 of which were violated.
We refined our initial $\mu$DD over several iterations, details of which we provide in Appendix \ref{apx:case-study-full-search}.
Across all explored models, there were thousands of \upaths and over a thousand model constraint violations.
Our guided refinement surpasses prior ad hoc reverse-engineering efforts \cite{babka2009investigating, wang2024peek, ertmerreverse, Zhao2022}, enabling us to uncover subtleties (with high confidence) in:

\vspace{1mm}
\noindent {\bf Address translation prefetchers.} Several studies have proposed address translation prefetching mechanisms, but little is known about how such prefetchers are actually implemented in real-world processors \cite{saulsbury2000recency,kandiraju2002going,bhattacharjee2010inter,vavouliotis2021morrigan,margaritov2021ptemagnet,vavouliotis2021exploiting}. Recent work suggests that underdocumented translation prefetching features may be at the core of unexplained performance anomalies in real-world workloads \cite{bertschi2022battling}.

Using \Counterpoint, we uncovered hardware in the Intel Haswell MMU that prefetches page table entries into its L1/L2 TLBs as well as PDE cache. Our analysis revealed three key aspects of the prefetcher’s implementation:

First, we identified prefetch trigger conditions. If a workload is feasible with an $\mu$DD that includes the prefetcher but infeasible in one without it, the workload must trigger prefetches. This helps us deduce that prefetching logic scans virtual page numbers in the load/store queue and is triggered by sequential accesses predicted to cross a page boundary—contradicting the common assumption that prefetches are triggered exclusively by TLB misses. For increasing virtual addresses, prefetching is triggered after consecutive accesses to cache lines 51 and 52 within a page; for decreasing addresses, the trigger occurs at cache lines 8 and 7. No other cache line pairs were observed to initiate prefetching.

Second, we found that the load/store queue logic responsible for virtual-page prefetching relies on the page table walker to resolve translation prefetch requests. In practice, this means that prefetches trigger the walker to inject additional load instructions into the CPU pipeline—the same way it injects loads for demand page table walks themselves (previously called “ghost” or “stuffed” loads \cite{Zhao2022, Lustig2016}). In some cases, the walker generates hundreds of such additional loads. This overturns the prevailing model in prior work, which assumed prefetches bypass the pipeline and enter the memory hierarchy directly, and therefore model prefetches with unrealistically low latency. It also implies that significantly more prefetches can be injected than previously believed.

Third, we found that prefetch-induced page table walks abort when they encounter a page table entry whose access (reference) bit is unset—unlike regular page table walks, which set this bit.
Consequently, the TLB prefetch does not complete. This behavior is logical: allowing a speculatively set access bit for an ineffective TLB prefetch could, in principle, lead to suboptimal paging decisions, and permitting TLB prefetches to set the access bit would also introduce additional microarchitectural complexity. Prefetch-induced page walks can still modify cache state, with potential performance and security implications. Some recent TLB prefetcher proposals allow prefetch-induced page walks to set the access bit and complete \cite{vavouliotis2021exploiting,vavouliotis2021morrigan}. While this behavior is architecturally permitted \cite{intel_ia32_sdm_2025}, we have not observed it on Haswell. \\

\vspace{1mm}
\noindent {\bf Page table walk merging.} Despite decades of research on TLBs and MMU caches, little is publicly known about how MMUs schedule page table walks. Using \Counterpoint, we discover that MMUs can {\it merge} multiple outstanding walks to the same virtual page into a single page table walk, which we capture by modeling MSHRs within our MMU $\mu$DD.

Historically, MMU MSHRs have not been modeled in address translation studies because their design involves subtleties beyond those of conventional cache MSHRs \cite{tuck2006scalable,kroft1998lockup,lee2009improving}. For instance, page sizes—and therefore virtual page numbers—are unknown until after translation \cite{cox2017efficient}, making MSHR lookup and allocation non-trivial. Further, distinct page table walks have unique rules for updating access and dirty bits, as well as for determining whether they are allowed to touch physical memory regions marked speculative versus non-speculative \cite{glew1997method}. These complexities make it far from obvious how walk merging can be safely implemented.

Our results show that MMU MSHRs are nonetheless critical for performance. For some workloads, page table walk merging reduces the number of distinct walks by nearly half. This finding underscores the importance of explicitly modeling MMU MSHRs in simulators used to evaluate address translation optimizations \cite{Bhattacharjee2018, bhattacharjee2019appendix, barr2010translation, Bhattacharjee2013largereach, Basu2013, Agbarya2020, babka2009investigating, Zhao2022,KanellopoulosVirtuosoMethodology,Bhattacharjee2017Translation-triggeredPrefetching}.

Finally, \Counterpoint reveals a surprising detail: the PDE cache is queried \emph{before} outstanding walks to the same virtual page are merged. Prior studies have not considered this interaction \cite{barr2010translation,Bhattacharjee2013largereach}. One might expect walk merging to reduce PDE cache lookups, easing port pressure, cutting bandwidth, and eliminating queuing delays. Instead, our $\mu$DD suggests that the PDE cache is looked up prior to MSHR allocation, likely to reduce latency via pipelining.  Importantly, \Counterpoint is able to do this because it enables discovery of not just individual hardware components, but also their relative placement within the pipeline.

\vspace{1mm}
\noindent {\bf Root-level MMU cache.} A large body of address translation research proposing hardware optimizations \cite{Bhattacharjee2013largereach,Bhattacharjee2018,yan2017hardware,8192494,margaritov2019prefetched,skarlatos2020elastic,Park2022} assumes the presence of a root-level MMU cache, yet some recent reverse-engineering studies have found no evidence of its existence \cite{van2017reverse, ertmerreverse}. \Counterpoint demonstrates its compatibility with all other address translation features identified in this paper for the workloads we analyze, giving architecture researchers confidence in including it in their models. When \emph{walk bypassing} is not modeled, several workloads become feasible only with a root-level MMU cache in the $\mu$DD. These workloads use 1GB pages which would stress a hypothetical PML4E cache (which is explicitly for 1GB pages), suggesting that for these workloads, the “missing” page table walker accesses could be explained by PML4E cache.

\vspace{1mm}
\noindent {\bf Aborted page table walks.} Recent research has studied page table walks under speculation in modern out-of-order processors \cite{lindsay2024understanding,Zhao2022,Gupta2021}. While prior work shows that x86-64 processors can abort in-flight page table walks in response to machine clears \cite{lindsay2024understanding,Ragab2021RageAttacks,Zhao2022}, the underlying implementation details remain poorly understood.

Using \Counterpoint, we find that aborted walks are entirely consistent with all our \HPC measurements and newly discovered features. Additionally, they appear to be triggered more frequently by workloads with high walker utilization.
While more detailed study is necessary to better understand how aborted page table walks are implemented, \Counterpoint suggests that page table walks can be aborted at any point—even before issuing a single memory access. This implies that aborted walks can still consume MMU and memory hierarchy resources, effectively imposing a hidden performance tax that should be explicitly modeled in simulation infrastructures for address translation \cite{KanellopoulosVirtuosoMethodology,Agbarya2020}.

\vspace{1mm}
\noindent {\bf Page table walk replays.} We observe that page table walks can complete without generating any memory accesses. This suggests that the core may include a mechanism allowing walks to finish without engaging the cache hierarchy. Prior work has shown a complex interplay between the hardware page table walker and microarchitectural structures that maintain memory consistency \cite{Zhao2022,pagewalk-coherence}. We hypothesize that these “missing” accesses occur because walks are replayed or handled by hidden internal address translation caching structures not reflected in the \PageWalkerLoads counter.
Understanding these structures more concretely would require implementing new \HPCs or access to proprietary RTL.
An alternative explanation is that the “missing” accesses do occur but are not counted because, unlike regular page-walker accesses, they are non-speculative. If we assume that aborted walks are replayed at micro-op retirement as non-speculative walks—as suggested in prior sources \cite{glew1997method,cordes_stack_overflow,pagewalk-coherence}—then the resulting $\mu$DD becomes feasible, but only once features such as TLB prefetching and miss merging are incorporated.

\subsection{\Counterpoint Performance Characterization} \label{sec:evaluation}

\begin{figure}
    \begin{subfigure}[b]{\linewidth}
        \centering
        \includegraphics[width=0.8\linewidth]{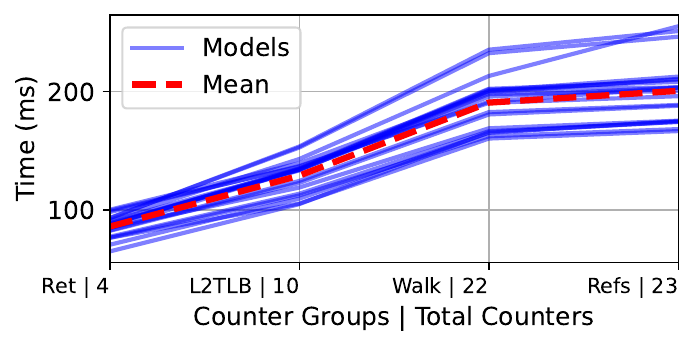}
        \caption{Evaluation time per observation per model.}
        \label{fig:feasibility-determination}
    \end{subfigure}
    \begin{subfigure}[b]{\linewidth}
    \centering
    \includegraphics[width=0.8\linewidth]{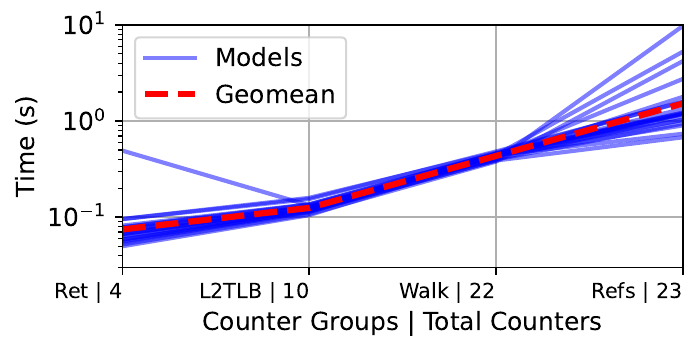}
    \caption{Constraint deduction time per model.}
    \label{fig:constraint-deduction}
    \end{subfigure}
    \caption{
    \Counterpoint performance (quantified for 20 representative $\mu$DDs) varies with models and scales with \HPCs.
    Blue lines represent individual models; groups of semantically related counters are added along x-axis. (a) Feasibility testing time scales linearly. (b) Constraint deduction time scales exponentially. The red lines represent (a) arithmetic, and (b) geometric means.
    }
\end{figure}

We evaluate \Counterpoint on a 24-core Intel Xeon E5-2680 v3 CPU running at 2.5GHz. For this analysis, we evaluate every observation against every constraint to produce a worst-case runtime, even though in practice only infeasible observations need to be checked. We parallelize the parts of \Counterpoint devoted to determining feasibility constraints. On average, \Counterpoint evaluates a model in 213 seconds, with the majority of time spent assessing individual model constraints. Determining model feasibility is much faster than checking each constraint, as model cones allow all constraints to be tested simultaneously. This makes it practical to evaluate models with large numbers of counters.

Figure \ref{fig:feasibility-determination} plots the time taken to determine observation feasibility as a function of the counters present in the model.
For the full suite of counters, \Counterpoint takes around 200 milliseconds to determine if an observation is feasible.
Empirically, the time taken scales approximately linearly as counter groups are added.
Additionally, observation feasibility testing is embarrassingly parallel, allowing large numbers of observations to be tested simultaneously.

Figure \ref{fig:constraint-deduction} shows how the time required to find the model constraints scales with the counter groups.
The logarithmic y-scale demonstrates that empirically the constraint deduction time scales exponentially as counter groups as added.
Despite this, \Counterpoint only takes between 800 milliseconds and 10 seconds to determine the set of constraints for models with all counters present.
Note that explicitly determining the model constraints is only used for providing feedback for model refinement; it is not a prerequisite for determining the feasibility of individual observations.

\section{Related Work} \label{sec:related-work}

Computer architects have recently used \HPCs to reverse-engineer specific microarchitectural features. \HPCs are used by \mbox{uops.info} \cite{abel2019uops} and \mbox{nanoBench} \cite{abel2020nanobench} to reverse-engineer micro-op performance and port assignment, as well as cache replacement policies.  Several studies have focused on reverse engineering the MMU \cite{babka2009investigating, wang2024peek, ertmerreverse, Zhao2022}, while Ragab et al \cite{Ragab2021RageAttacks} use \HPCs to characterize the security implications of machine clears.
Binoculars \cite{Zhao2022} use \HPCs to characterize page table walker contention.
With \udd models, \Counterpoint offers a more general-purpose approach.

Multiplexing noise is a well-studied problem \cite{Lv2018,AzimiOnlineCounters,Banerjee2021}.
Azimi et al \cite{azimi2009enhancing} quantify multiplexing noise for a range of workloads.
CounterMiner \cite{Lv2018} replacing outliers with interpolated values.
BayesPerf \cite{Banerjee2021} reduces noise by exploiting \emph{known} statistical relationships between counter values.
\Counterpoint infers correlations to reduce noise impact.

Interpreting \HPC values correctly remains challenging.
Vendors provide explicit metrics that convert \HPC values to standard metrics (\eg, CPI, MPKI, hit rates, \etc), but not all \HPCs are used.
The Counter Inspection Toolkit \cite{danalis2017counter} and related work \cite{barry2024automated,barry2021effortless} correlate counters with individual microbenchmarks to define new metrics.
Top Down Methodology \cite{Yasin2014} employs metrics and thresholds to enable application developers to identify performance bottlenecks.
Unlike bespoke approaches, \Counterpoint $\mu$DDs capture both \HPC semantics and microarchitectural features by construction.

Formal modeling for microarchitectures has recently been used for memory consistency \cite{Manerkar2018,Lustig2015,Lustig2016,Hossain2020,Norman2023PipeSynth:Consistency, Hsiao2021SynthesizingImplementations}, cache coherence \cite{pong1993verification,pong1995new,oswald2018protogen,oswald2020hieragen,oswald2022heterogen,dave2005automatic}, and security \cite{Trippel2018,Hsiao2024}.
CheckSuite and related tools \cite{Manerkar2018,Lustig2015,Lustig2016,Hossain2020,Norman2023PipeSynth:Consistency, Hsiao2021SynthesizingImplementations,Trippel2018,Hsiao2024} describe microarchitectural executions using $\mu$spec models featuring \upaths and inter- and intra-\upath dependencies.
\Counterpoint's $\mu$DD formalism is compatible with these approaches.

\section{Conclusion \& Future Work}\label{sec:conclusion}
We presented \Counterpoint, a framework that transforms large \HPC datasets into accurate, high-quality microarchitectural models. By encoding an expert’s mental model as a $\mu$DD and automatically generating model constraints, \Counterpoint eliminates the tedium and errors of manual derivation. At the same time, \Counterpoint processes noisy \HPC measurements into reliable, high-confidence ranges, bridging intuition with data-driven analysis. \Counterpoint accelerates and sharpens the modeling of complex architectures, freeing experts to focus on insight rather than bookkeeping, and making advanced microarchitectural modeling faster and more insightful. By accelerating the productivity of these influential experts, insights extracted by \Counterpoint's have the potential to shape the broader field of computing.

While this first paper demonstrates the promise of \Counterpoint, several productive directions remain for future work. For example, \Counterpoint could potentially be used to reverse-engineer not only microarchitectural details but also the semantics of undocumented HECs (see our work on reverse-engineering the semantics of the \PageWalkerLoads counter for page table walk replays). Establishing this, however, would require a detailed study beyond the scope of this work, which we leave for future work. Additionally, our current study evaluates the benefit of \Counterpoint on CPUs; exploring the utility of \Counterpoint to hardware accelerators would broaden its applicability. Finally, this paper presents a first proof-of-concept study of the key ideas and principles behind \Counterpoint. Substantial work remains to extend it into a robust, system-wide modeling framework, including support for multiple cores, multiple sockets, hyperthreading, kernel-level activity, and much more.

\section{Acknowledgments}\label{sec:acks}
We thank the anonymous reviewers, Neil Zhao, Michael Wu, and Grace Jia for their valuable feedback on drafts of this paper. We also thank Manolis Zampetakis for his insights on our constraint deduction algorithm and for highlighting potential directions for future optimizations. We appreciate the technical discussions with Lieven Eeckhout, Andrew Milas, and Matt Sinclair. This work was supported in part by National Science Foundation awards 2236855, 2112562, and 2047220. We thank Intel for supporting the IISWC 2024 paper that preceded and motivated this work.

\bibliographystyle{ACM-Reference-Format}
\bibliography{references}

\clearpage
\appendix

\section{Linear Program Formulation} \label{apx:linear-program}

We construct and solve the following linear program to determine the feasibility of microarchitectural observations against a $\mu$DD model:
\begin{equation} \tag{LP} \label{eqn:feasibility-lp}
\left[
\begin{aligned}
    &\vec{v} \in \mathbb{R}_{+}^N    \hfill                                                                       \textcolor{gray}{\textrm{ (Counter variables)}}&\\
    &\forall p \in \setofpaths{D}\ .\ f(p) \in \mathbb{R}_{+}    \hfill                                                                       \textcolor{gray}{\textrm{ (Flow variables)}}&\\
    &\vec{v} = \sum_{p \in \setofpaths{D}} \sig{p} \cdot f(p)    \hfill                                           \textcolor{gray}{\textrm{ (Counter flow equation)}}&\\
    &\forall i \in n \ .\  |\vec{e}_i \cdot (\vec{v} - \bar{Y})| \leq \sqrt{\lambda_i \chi_{d,1-\alpha}^{2}} \hfill&\\
    & \hfill \textcolor{gray}{\textrm{ (Counter confidence region encoding)}}& \\
\end{aligned}
\right]
\end{equation}
Each path through the $\mu$DD is enumerated by breadth-first search.
Variables are instantiated for the the true counter values $\vec{v}$ and the flow $f(p)$ down each \upath.
The variables are constrained to be non-negative.
Counter and flow variables are related by the \ref{eqn:cfe}, which implicitly describes the model cone.

The confidence ellipsoid cannot be directly encoded in the linear program as it is a quadratic form.
Instead, the bounding box is constructed - aligning edges to the principle axes of the ellipsoid (Figure \ref{fig:confidence-region-bounding-box}).
The directions of the principle axes of the confidence ellipsoid are determined by the normalized eigenvectors $\vec{e_1},...,\vec{e_n}$ of the covariance matrix.
The half-length of the $i$th axis is given by $\sqrt{\lambda_i \chi_{N, 1-\alpha}^2}$, where $\lambda_i$ is the $i$th eigenvalue.
Figure \ref{fig:confidence-region-bounding-box} graphically depicts the construction for systems with two counters.

\section{Hardware Performance Counter Events} \label{apx:performance_counter_events}

\newcommand{\TT}{\textbf{T}}

\newcommand{\TTp}{\textbf{T'}}
\newcommand{\PWL}{\textbf{pwl}}
\newcommand{\MUR}{\textbf{mur}}

\begin{table}[]
\centering
\caption{Hardware event counters used in paper.
\textbf{Grp} is our event group classification, \textbf{This Paper} is the \HPC name used in this paper, and \textbf{Full Event Name Suffix} is the suffix of the full event named described in the Linux \texttt{perf} event database \cite{linux-pmu-events}.
All events other than Refs are parameterized by access type $\TT \in \{\texttt{load},\texttt{store}\}$.
All Walk and STLB group events have full event names prefixed by \texttt{stlb\_\TT\_misses}.
All Refs events have a full event name prefixed by \texttt{page\_walker\_loads}.
All Ret events have a full event name prefixed by \texttt{mem\_uops\_retired}.}.
\label{tab:hardware-event-counters}
\begin{tabular}{|l|l|l|}
\hline
\textbf{Grp} & \textbf{This Paper} & \textbf{Full Event Name Suffix} \\ \hline

\multirow{6}{*}{\begin{tabular}[c]{@{}l@{}}Walk\\ (12)\end{tabular}} 
 & \texttt{\TT.causes\_walk}        
 & \texttt{miss\_causes\_a\_walk} \\ \cline{2-3}

 & \texttt{\TT.walk\_done\_4k}      
 & \texttt{walk\_completed\_4k} \\ \cline{2-3}

 & \texttt{\TT.walk\_done\_2m}      
 & \texttt{walk\_completed\_2m\_4m} \\ \cline{2-3}

 & \texttt{\TT.walk\_done\_1g}      
 & \texttt{walk\_completed\_1g} \\ \cline{2-3}

 & \texttt{\TT.walk\_done}          
 & \texttt{walk\_completed} \\ \cline{2-3}

 & \texttt{\TT.pde\$\_miss}         
 & \texttt{pde\_cache\_miss} \\ \hline

\multirow{4}{*}{\begin{tabular}[c]{@{}l@{}}Refs\\ (4)\end{tabular}} 
 & \texttt{walk\_ref.l1}            
 & \texttt{dtlb\_l1} \\ \cline{2-3}

 & \texttt{walk\_ref.l2}            
 & \texttt{dtlb\_l2} \\ \cline{2-3}

 & \texttt{walk\_ref.l3}            
 & \texttt{dtlb\_l3} \\ \cline{2-3}

 & \texttt{walk\_ref.mem}           
 & \texttt{memory} \\ \hline

\multirow{2}{*}{\begin{tabular}[c]{@{}l@{}}Ret\\ (4)\end{tabular}} 
 & \texttt{\TT.ret\_stlb\_miss}     
 & \texttt{stlb\_miss\_\TT s} \\ \cline{2-3}

 & \texttt{\TT.ret}                
 & \texttt{all\_\TT s} \\ \hline

\multirow{3}{*}{\begin{tabular}[c]{@{}l@{}}STLB\\ (6)\end{tabular}} 
 & \texttt{\TT.stlb\_hit\_4k}       
 & \texttt{stlb\_hit\_4k} \\ \cline{2-3}

 & \texttt{\TT.stlb\_hit\_2m}       
 & \texttt{stlb\_hit\_2m} \\ \cline{2-3}

 & \texttt{\TT.stlb\_hit}           
 & \texttt{stlb\_hit} \\ \hline

\end{tabular}
\end{table}

Table \ref{tab:hardware-event-counters} lists the hardware event counters and their group classification used within this paper.

\section{Case Study: Full Search Procedure} \label{apx:case-study-full-search}

CounterPoint is an automated approach and tool for refuting and refining $\mu$DDs given a set of programs. The set of candidate $\mu$DDs depends on the aspects of the microarchitecture the user wishes to capture, and the set of feasible $\mu$DDs depends on the programs being recorded.
Suites of programs that exhibit diverse microarchitectural behaviors enable models of greater fidelity. In exploring address translation on Intel Haswell, we created and tested dozens of $\mu$DDs, and we continue to refine our models.

\subsection{Initial model search} \label{apx:case-study-initial-search}

We identified many microarchitectural features in the initial phase of our search.
We describe our initial model assumptions, corresponding to model \texttt{m0} in Table \ref{tab:search-initial}, in Section \ref{sec:case_study}.
Note that we assume there are two fundamental micro-op types (load and store), and that only micro-ops that obtain a valid translation are allowed to retire.

\begin{figure}
    \centering
    \includegraphics[width=\linewidth]{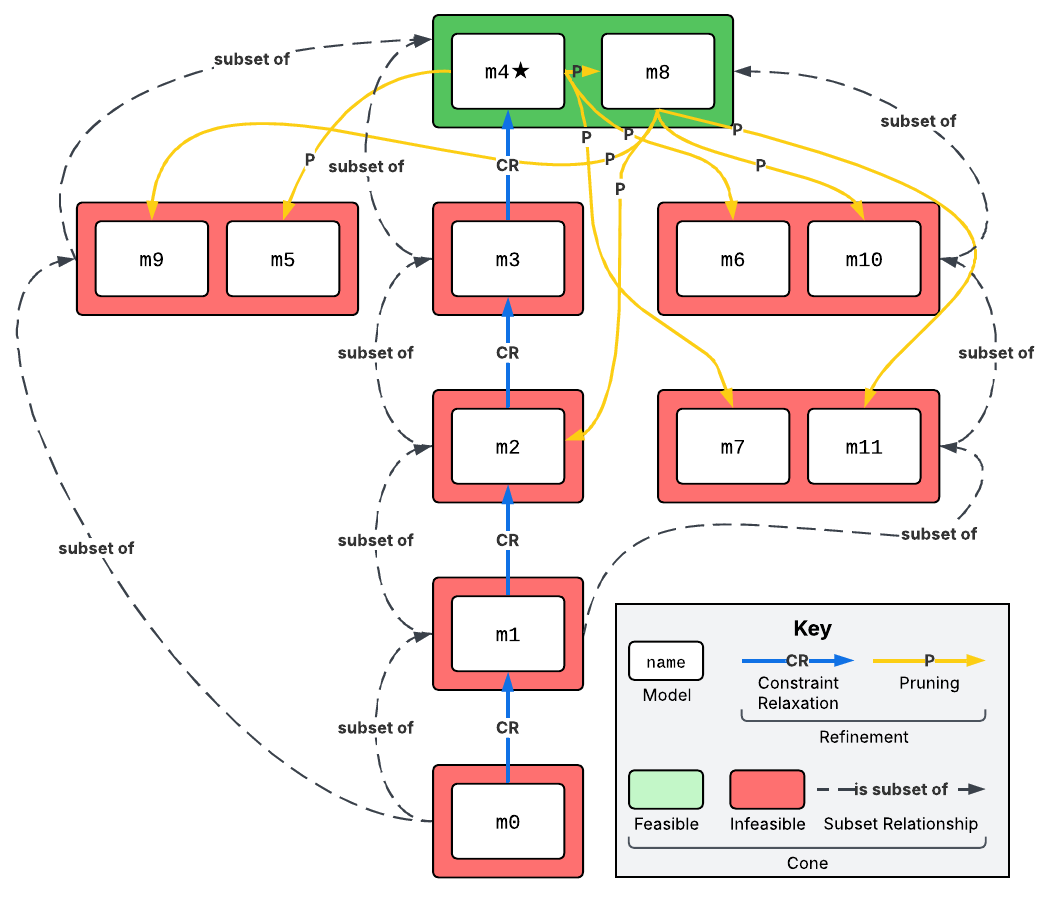}
    \caption{Models and model cones obtained by our initial search procedure. Even exploring a small number of models and features yields elaborate relationships among model cones.}
    \label{fig:search-initial}
\end{figure}

\DeclareRobustCommand{\yesyes}{\raisebox{0pt}[0pt][0pt]{\textcolor{green!60!black}{\ding{51}}}}
\DeclareRobustCommand{\nono}{\raisebox{0pt}[0pt][0pt]{\textcolor{red}{\ding{55}}}}

\begin{table}[]
\centering
\caption{$\mu$DDs explored in the initial search. Models are identified by their name (left column), features (middle columns), and number of infeasible observations (right column).}
\label{tab:search-initial}

\begin{tabular}{|r|ccccc|l|}
\hline
\textbf{} & \textbf{\begin{tabular}[c]{@{}c@{}}Tlb\\ Pf*\end{tabular}} & \textbf{\begin{tabular}[c]{@{}c@{}}Early\\ Psc\end{tabular}} & \textbf{\begin{tabular}[c]{@{}c@{}}Merg-\\ ing\end{tabular}} & \textbf{\begin{tabular}[c]{@{}c@{}}Pml4e\\ Cache\end{tabular}} & \textbf{\begin{tabular}[c]{@{}c@{}}Walk\\ Bypass\end{tabular}} & \textbf{\begin{tabular}[c]{@{}l@{}}\# \\ Inf.\end{tabular}} \\ \hline
m0 & \nono & \nono & \nono & \nono & \nono & 209 \\
m1 & \yesyes & \nono & \nono & \nono & \nono & 204 \\
m2 & \yesyes & \yesyes & \yesyes & \nono & \nono & 91 \\
m3 & \yesyes & \yesyes & \yesyes & \yesyes & \nono & 56 \\
$\bigstar$ \textbf{m4} & \textbf{\yesyes} & \textbf{\yesyes} & \textbf{\yesyes} & \textbf{\yesyes} & \textbf{\yesyes} & \textbf{0} \\
m5 & \nono & \yesyes & \yesyes & \yesyes & \yesyes & 5 \\
m6 & \yesyes & \nono & \yesyes & \yesyes & \yesyes & 142 \\
m7 & \yesyes & \yesyes & \nono & \yesyes & \yesyes & 143 \\
m8 & \yesyes & \yesyes & \yesyes & \nono & \yesyes & 0 \\
m9 & \nono & \yesyes & \yesyes & \nono & \yesyes & 5 \\
m10 & \yesyes & \nono & \yesyes & \nono & \yesyes & 142 \\
m11 & \yesyes & \yesyes & \nono & \nono & \yesyes & 143 \\ \hline
\end{tabular}

\raggedright {\small *TLB Prefetching.}
\end{table}

\begin{table}[]
\caption{Microarchitecture features (initial model search).}
\label{tab:features-initial}
\begin{tabular}{|r|l|}
\hline
\textbf{Feature}                                           & \textbf{Description}                                                                                               \\ \hline
\begin{tabular}[c]{@{}r@{}}TLB \\ Prefetching\end{tabular} & \begin{tabular}[c]{@{}l@{}}Prefetches form an additional kind of \\ translation requests\end{tabular}              \\ \hline
Early PSC                                                  & \begin{tabular}[c]{@{}l@{}}Paging structure caches are looked up \\ before starting a walk\end{tabular}            \\ \hline
Merging                                                    & \begin{tabular}[c]{@{}l@{}}Page table walks can be merged by an \\ L2TLB MSHR\end{tabular}                         \\ \hline
\begin{tabular}[c]{@{}r@{}}PML4E \\ Cache\end{tabular}     & \begin{tabular}[c]{@{}l@{}}There exists paging structure cache for\\ root (PML4E) level of page table\end{tabular} \\ \hline
\begin{tabular}[c]{@{}r@{}}Walk \\ Bypass\end{tabular}     & \begin{tabular}[c]{@{}l@{}}Walks can complete without making \\ visible memory access\end{tabular}                 \\ \hline
\end{tabular}
\end{table}

We explore a wide range of models (Table \ref{tab:search-initial}) using the search procedure outlined in Section \ref{sec:guided-model-exploration}.
Figure \ref{fig:search-initial} shows the explored search space.
Each white box indicates a particular $\mu$DD.
Edges connect models depending on whether they were derived through constraint relaxation (blue edges) or by feature pruning (yellow edges).
Each $\mu$DD is associated with a model cone, which is either feasible (green) or infeasible (red).
Multiple $\mu$DDs might produce the same model cone, as illustrated by a model cone box containing more than one model.

Tables \ref{tab:search-initial}, \ref{tab:features-initial} list the features associated with each model that we explore.
Models \texttt{m4} and \texttt{m8} are identified as feasible. For the purposes of this search methodology, we consider \texttt{m4} as our model because experts assume its presence in typical address translation research studies. A separate search trajectory could be invoked using \texttt{m8} as the starting point for the continued trajectory.

\subsection{TLB prefetch trigger conditions} \label{apx:case-study-prefetch-trigger}

We further refine the TLB trigger conditions by removing the abstract \emph{prefetch} translation request type and instead associate TLB prefetches directly with their triggering \uop.

\begin{table}[]
\caption{$\mu$DDs explored in TLB prefetching analysis. All $\mu$DDs are derivatives of \texttt{m4}.}
\label{tab:search-prefetch}

\begin{tabular}{r|ccccc|l}
\cline{2-6}
\multicolumn{1}{l|}{} & \multicolumn{5}{c|}{\textbf{TLB Prefetch Trigger Conditions}} &  \\ \hline
\multicolumn{1}{|r|}{\textbf{}} & \textbf{Spec.} & \textbf{Load} & \textbf{Store} & \textbf{\begin{tabular}[c]{@{}c@{}}Dtlb\\ Miss\end{tabular}} & \textbf{\begin{tabular}[c]{@{}c@{}}Stlb\\ Miss\end{tabular}} & \multicolumn{1}{l|}{\textbf{\begin{tabular}[c]{@{}l@{}}\#\\ Inf.\end{tabular}}} \\ \hline
\multicolumn{1}{|r|}{$\bigstar$ \textbf{t0}} & \textbf{\yesyes} & \textbf{\yesyes} & \textbf{\nono} & \textbf{\nono} & \textbf{\nono} & \multicolumn{1}{l|}{\textbf{0}} \\
\multicolumn{1}{|r|}{t1} & \yesyes & \yesyes & \nono & \yesyes & \nono & \multicolumn{1}{l|}{0} \\
\multicolumn{1}{|r|}{t2} & \yesyes & \yesyes & \nono & \nono & \yesyes & \multicolumn{1}{l|}{0} \\
\multicolumn{1}{|r|}{t3} & \yesyes & \nono & \yesyes & \nono & \nono & \multicolumn{1}{l|}{0} \\
\multicolumn{1}{|r|}{t4} & \yesyes & \nono & \yesyes & \yesyes & \nono & \multicolumn{1}{l|}{0} \\
\multicolumn{1}{|r|}{t5} & \yesyes & \nono & \yesyes & \nono & \yesyes & \multicolumn{1}{l|}{0} \\
\multicolumn{1}{|r|}{t6} & \yesyes & \yesyes & \yesyes & \nono & \nono & \multicolumn{1}{l|}{0} \\
\multicolumn{1}{|r|}{t7} & \yesyes & \yesyes & \yesyes & \yesyes & \nono & \multicolumn{1}{l|}{0} \\
\multicolumn{1}{|r|}{t8} & \yesyes & \yesyes & \yesyes & \nono & \yesyes & \multicolumn{1}{l|}{0} \\
\multicolumn{1}{|r|}{t9} & \nono & \yesyes & \nono & \nono & \nono & \multicolumn{1}{l|}{0} \\
\multicolumn{1}{|r|}{t10} & \nono & \yesyes & \nono & \yesyes & \nono & \multicolumn{1}{l|}{4} \\
\multicolumn{1}{|r|}{t11} & \nono & \yesyes & \nono & \nono & \yesyes & \multicolumn{1}{l|}{4} \\
\multicolumn{1}{|r|}{t12} & \nono & \nono & \yesyes & \nono & \nono & \multicolumn{1}{l|}{0} \\
\multicolumn{1}{|r|}{t13} & \nono & \nono & \yesyes & \yesyes & \nono & \multicolumn{1}{l|}{4} \\
\multicolumn{1}{|r|}{t14} & \nono & \nono & \yesyes & \nono & \yesyes & \multicolumn{1}{l|}{4} \\
\multicolumn{1}{|r|}{t15} & \nono & \yesyes & \yesyes & \nono & \nono & \multicolumn{1}{l|}{0} \\
\multicolumn{1}{|r|}{t16} & \nono & \yesyes & \yesyes & \yesyes & \nono & \multicolumn{1}{l|}{3} \\
\multicolumn{1}{|r|}{t17} & \nono & \yesyes & \yesyes & \nono & \yesyes & \multicolumn{1}{l|}{4} \\ \hline
\end{tabular}

\end{table}

\begin{table}[]
\caption{Candidate TLB prefetcher trigger conditions.}
\label{tab:tlbpf-triggers}
\begin{tabular}{|r|l|}
\hline
\textbf{Condition}                                     & \textbf{Description}                                                                                                \\ \hline
Spec                                                 & \begin{tabular}[c]{@{}l@{}}Can be triggered by purely speculative \\ micro-ops (versus only retiring).\end{tabular} \\ \hline
Load                                                 & Can be triggered by load micro-ops                                                                                  \\ \hline
Store                                                & Can be triggered by store micro-ops                                                                                 \\ \hline
\begin{tabular}[c]{@{}r@{}}L1TLB\\ Miss\end{tabular} & \begin{tabular}[c]{@{}l@{}}Demand L1TLB misses can cause \\ prefetcher to inject page table walk.\end{tabular}      \\ \hline
\begin{tabular}[c]{@{}r@{}}L2TLB\\ Miss\end{tabular} & \begin{tabular}[c]{@{}l@{}}Demand L2TLB misses can cause \\ prefetcher to inject page table walk.\end{tabular}      \\ \hline
\end{tabular}
\end{table}

We generate 18 separate models (Table \ref{tab:search-prefetch}), each a variant of the \texttt{m4} model but with different TLB prefetch trigger conditions.
Tables \ref{tab:search-prefetch}, \ref{tab:tlbpf-triggers} list the models and trigger conditions.

Feasibility analysis (Table \ref{tab:search-prefetch}) reveals that all $\mu$DDs that allow TLB prefetches to be triggered by speculative micro-ops are feasible.
If prefetching is restricted to only non-speculative micro-ops, then the TLB prefetcher can only be triggered before DTLB lookup (prefetches cannot be triggered by the DTLB or STLB miss stream).

We make further insights based on the following heuristic.
All workloads that require TLB prefetching are specific instances of the linear access microbenchmark.
This microbenchmark consists of an infinite \texttt{while} loop of memory accesses that is terminated after 10 minutes.
This is an extremely simple control flow pattern that the branch predictor should be able to learn perfectly.
For this reason, we assume that all micro-ops in the workload eventually retire (i.e., that \HPC increments for micro-ops that do not retire are absorbed by the confidence region).
Therefore, we can assume that our microbenchmark consists solely of retiring micro-ops, and we can use the feasibility results for models with non-speculative TLB prefetching triggers to determine the overall feature set, \emph{provided we restrict our analysis to the linear-access microbenchmark}.
Analysis of these models with this microbenchmark reveals that TLB prefetching \emph{must} be triggered \emph{prior} to DTLB lookups (\eg, in the load-store queue).

Furthermore, no instances of our microbenchmark with a store-only access pattern trigger TLB prefetching (i.e., the sequential access microbenchmark with stores does not violate any constraint that is relaxed by TLB prefetching).
This leads us to believe that only load micro-ops can trigger TLB prefetches.

For these reasons, and for the purpose of demonstration, we believe $\mu$DD \texttt{t0} to be a representative model.
We were unable to determine whether speculative non-retired micro-ops can trigger the TLB prefetcher; therefore, we conservatively assume that all load micro-ops (including purely speculative load micro-ops) can trigger TLB prefetching.
We leave determining if wrong-path speculative load micro-ops can trigger the TLB prefetcher for future work.

\subsection{Aborts as alternative to walk bypassing} \label{apx:case-study-aborts-vs-bypassing}

\begin{table}[]
\caption{Models with different abort points.}
\label{tab:abort-point-udds}
\begin{tabular}{l|cccc|l}
\cline{2-5}
                                & \multicolumn{4}{c|}{\textbf{Translation Request Abort Point}}                                                                                                                                                                                                   &                                                                                 \\ \hline
\multicolumn{1}{|l|}{\textbf{}} & \textbf{\begin{tabular}[c]{@{}c@{}}During\\ Walk\end{tabular}} & \textbf{\begin{tabular}[c]{@{}c@{}}After\\ PSC\end{tabular}} & \textbf{\begin{tabular}[c]{@{}c@{}}After\\ L2TLB\end{tabular}} & \textbf{\begin{tabular}[c]{@{}c@{}}After\\ L1TLB\end{tabular}} & \multicolumn{1}{l|}{\textbf{\begin{tabular}[c]{@{}l@{}}\#\\ Inf.\end{tabular}}} \\ \hline
\multicolumn{1}{|l|}{a0}        & \yesyes                                                        & \nono                                                        & \nono                                                          & \nono                                                          & \multicolumn{1}{l|}{37}                                                         \\
\multicolumn{1}{|l|}{a1}        & \yesyes                                                        & \yesyes                                                      & \nono                                                          & \nono                                                          & \multicolumn{1}{l|}{37}                                                         \\
\multicolumn{1}{|l|}{a2}        & \yesyes                                                        & \yesyes                                                      & \yesyes                                                        & \nono                                                          & \multicolumn{1}{l|}{37}                                                         \\
\multicolumn{1}{|l|}{a3}        & \yesyes                                                        & \yesyes                                                      & \yesyes                                                        & \yesyes                                                        & \multicolumn{1}{l|}{37}                                                         \\ \hline
\end{tabular}
\end{table}

We were interested in further exploring mechanisms for \emph{walk bypassing}.
We consider page table walk \emph{aborts}, as described by Zhao et al. \cite{Zhao2022}, as an alternative to our proposed walk-bypassing feature.
Using \texttt{t0} as an example starting point, we replaced \emph{walk bypassing} with translation request aborts at four locations within the MMU pipeline (Table \ref{tab:abort-point-udds}).
None of the resulting models were feasible - not even the most aggressive, which allows aborts at all pipeline stages.
This indicates that, if model \texttt{t0} is accurate, translation aborts alone are insufficient to explain the "missing" memory accesses accounted for by walk bypassing.

\subsection{Page table walk replays} \label{apx:case-study-walk-replays}

We consider page table walk \emph{replays} as an alternative to walk bypassing.
An Intel patent \cite{glew1997method} describes a mechanism to implement page table walks on an out-of-order processor with speculative execution:
walks can be performed for speculative instructions, however, under certain conditions (e.g., invalid PTEs \cite{cordes_stack_overflow}, unset Accessed/Dirty bits \cite{glew1997method}, and memory ordering conflicts \cite{pagewalk-coherence}), the page table walk must be replayed should the \uop reach the head of the ROB (e.g., the \uop is not squashed).

Choosing \texttt{t0} as an example starting point, we replaced the walk-bypassing feature with the walk-replay feature described above.
The resulting model was found to be feasible.
The model relies on an assumption that memory references made by a \emph{replayed} walk are not recorded by any of the \PageWalkerLoads counters.
We justify this assumption because, unlike regular page walker accesses, replay accesses have special attributes, and accesses with such attributes may not be captured by \PageWalkerLoads.
In particular, replay walk accesses are \emph{non-speculative}, which enables them to access \emph{uncacheable} memory locations that regular speculative walks cannot access \cite{intel_ia32_sdm_2025}.

The walk replay mechanism requires that speculative walks can be aborted, so we include a walk abort feature.
We find that removing other features identified in this work (such as miss-merging) makes the resulting model infeasible.
This highlights that \Counterpoint's holistic modeling strategy can discover rich microarchitectural interactions that prior work, which considers features in isolation, does not.

\end{document}